\documentclass[twocolumn]{aastex62}

\usepackage{natbib}
\usepackage{amsmath}
\usepackage{amssymb}
\usepackage{amsfonts}
\usepackage{graphicx}
\usepackage{hyperref}

\newcommand{\gwbns}{GW170817A}
\newcommand{\grbbns}{GRB 170817A}
\newcommand{\afterglowpy}{{\tt afterglowpy}}
\newcommand{\boxfit}{{\tt BoxFit}}
\newcommand{\python}{{\tt Python}}
\newcommand{\emcee}{{\tt emcee}}

\newcommand{\hubble}{{\em Hubble Space Telescope}}
\newcommand{\planck}{{\em Planck}}

\newcommand{\swiftXRT}{{\em Swift-XRT}}

\newcommand{\dd}{\ensuremath{\mathrm{d}}}
\newcommand{\Mej}{\ensuremath{M_{\mathrm{ej}}}}
\newcommand{\tobs}{\ensuremath{t_{\mathrm{obs}}}}
\newcommand{\tW}{\ensuremath{t_{\mathrm{w}}}}
\newcommand{\tNR}{\ensuremath{t_{\mathrm{NR}}}}
\newcommand{\tb}{\ensuremath{t_{\mathrm{b}}}}

\newcommand{\tbin}{\ensuremath{t_{\mathrm{b, in}}}}
\newcommand{\tbout}{\ensuremath{t_{\mathrm{b, out}}}}
\newcommand{\tdec}{\ensuremath{t_{\mathrm{dec}}}}
\newcommand{\tumin}{\ensuremath{t_{\mathrm{min}}}}
\newcommand{\nuobs}{\ensuremath{\nu_{\mathrm{obs}}}}
\newcommand{\thobs}{\ensuremath{\theta_{\mathrm{obs}}}}
\newcommand{\phobs}{\ensuremath{\phi_{\mathrm{obs}}}}
\newcommand{\thW}{\ensuremath{\theta_{\mathrm{w}}}}
\newcommand{\thC}{\ensuremath{\theta_{\mathrm{c}}}}
\newcommand{\epse}{\ensuremath{\varepsilon_{\mathrm{e}}}}

\newcommand{\epsB}{\ensuremath{\varepsilon_{\mathrm{B}}}}
\newcommand{\xiN}{\ensuremath{\xi_N}}

\newcommand{\dL}{\ensuremath{d_{\mathrm{L}}}}

\newcommand{\Mp}{\ensuremath{m_{\mathrm{p}}}}

\newcommand{\Eiso}{\ensuremath{E_{\mathrm{iso}}}}
\newcommand{\Etot}{\ensuremath{E_{\mathrm{tot}}}}
\newcommand{\geff}{\ensuremath{g_{\mathrm{eff}}}}
\newcommand{\theff}{\ensuremath{\theta_{\mathrm{eff}}}}
\newcommand{\umin}{\ensuremath{u_{\mathrm{min}}}}
\newcommand{\umax}{\ensuremath{u_{\mathrm{max}}}}

\newcommand{\gmax}{\ensuremath{\gamma_{\mathrm{max}}}}

\newcommand{\som}{\ensuremath{s_{\Omega}}}

\begin{document}

\title{Gamma-Ray Burst Afterglows In The Multi-Messenger Era: Numerical Models and Closure Relations}

\correspondingauthor{Geoffrey Ryan}
\email{gsryan@umd.edu}

\author[0000-0001-9068-7157]{Geoffrey Ryan}
\altaffiliation{JSI Fellow}
\affil{Joint Space-Science Institute, University of Maryland, College Park, MD 20742, USA}

\author{Hendrik van Eerten}
\affil{Department of Physics, University of Bath, Claverton Down, Bath BA2 7AY, United Kingdom}
\author{Luigi Piro}
\affil{INAF, Istituto di Astrofisica e Planetologia Spaziali, via Fosso del Cavaliere 100, 00133 Rome, Italy}
\author{Eleonora Troja}
\affil{Department of Astronomy, University of Maryland, College Park, MD 20742-4111, USA}
\affil{Astrophysics Science Division, NASA Goddard Space Flight Center, 8800 Greenbelt Rd, Greenbelt, MD 20771, USA}

\begin{abstract}
	Gamma-ray bursts (GRBs) associated with gravitational wave events are, and will likely continue to be, viewed at a larger inclination than GRBs without gravitational wave detections.  As demonstrated by the afterglow of \gwbns{}, this requires an extension of the common GRB afterglow models which typically assume emission from an on-axis top hat jet.  We present a characterization of the afterglows arising from structured jets, providing a framework covering both successful and choked jets.  We compute new closure relations for structured jets and compare with the established relations for energy injection and refreshed shock models.  The temporal slope before the jet break is found to be a simple function of the ratio between the viewing angle and effective opening angle of the jet.  A numerical model to calculate synthetic light curves and spectra is publicly available as the open source \python{} package \afterglowpy{}.

\end{abstract}

\section{Introduction}

The binary neutron star merger event \gwbns{}, followed quickly by the short gamma-ray burst \grbbns{}, provided a new view on gamma-ray burst (GRB) afterglows \citep{Abbott:2017ab, Abbott:2017ac}.  Unlike typical afterglows, which begin bright and decay from detection with a ~week timescale, \grbbns{}'s non-thermal emission was undetectable until the first observation of X-rays nine days after the GRB \citep{Troja:2017aa}.  At this point the afterglow began steadily increasing in brightness at all wavelengths for $\sim160$ days \citep{Haggard:2017aa, Hallinan:2017aa, DAvanzo:2018aa,  Lyman:2018aa, Margutti:2018aa, Mooley:2018aa, Troja:2018aa, Troja:2019ab}.  Very long baseline (VLBI) radio imagery over this period identified a radio core with an apparent superluminal motion, indicating the emitting surface was moving at relativistically at an oblique angle towards the Earth \citep{Mooley:2018ab}. The emission peaked at 164 days after the burst and proceeded to sharply decay at a rate $t^{-2.2}$ commensurate with other GRB afterglows \citep{Alexander:2018aa, Fong:2019aa, Lamb:2019aa, Troja:2019ab}.

The lack of early emission, slow rising light curve, apparent motion of the radio centroid, and sharp decline post-peak are all consistent with emission from a \emph{structured jet}, a collimated blast wave with a non-trivial angular distribution of energy, viewed a moderate angle away from the jet axis \citep{Lamb:2017aa, Alexander:2018aa, Hotokezaka:2018aa, Wu:2018aa, Xie:2018aa,  Ghirlanda:2019aa, Fong:2019aa, Lamb:2019aa, Troja:2019aa}.  The combined effects of a moderate viewing angle and angular structure in the jet produce a light curve significantly different from the standard on-axis uniform ``top-hat'' jet used extensively in GRB afterglow analysis. Being sources at cosmological distances, GRBs must typically be observed nearly on-axis or at least within the original opening angle of the jet \citep{Ryan:2015aa}. At these small viewing angles lateral structure plays a sub-dominant role and the ubiquitous top hat jet has been a sufficient model for most studies.  Since the gravitational wave signal of a binary neutron star merger is nearly isotropic, it is expected the majority of future GW-GRBs will be viewed at significant inclination and may have similarly peculiar light curves as \grbbns{}.

A solid theoretical understanding of structured jet afterglows, including viewing angle effects, will be required to make the most use of future GW-GRB observations, the \grbbns{} dataset, and re-analysis of archival short GRBs \citep{Troja:2018ab, Troja:2019aa}. Standard tools for GRB analysis include closure relations, equations relating the temporal and spectral power law slopes of GRB afterglow light curves (e.g. \citet{Granot:2002aa, Racusin:2009aa}), jet breaks, achromatic breaks in the light curve related to the opening angle of the jet, and full light curve modeling.  To this end, we have calculated generalized closure relations for GRB afterglows including explicit viewing angle and jet structure dependence, putting these effects into a similar framework as standard energy injection models \citep{Zhang:2006aa}.
 We have also developed the computational tool \afterglowpy{}: a public, open source \python{} package for on-the-fly computation of structured jet afterglows with arbitrary viewing angle.

Jet structure and viewing angle have a long history as potential explanations for temporal behaviors and breaks in GRB afterglows.  \citet{Meszaros:1998aa} first considered anisotropic models and noted they may present as ``orphan afterglows'' without a prompt GRB signal if viewed sufficiently off-axis.  Detailed theoretical calculations of structured jet afterglow light curves concluded several facts: on-axis structured light curves resemble those from top hat jets, off-axis structured light curves show an achromatic break at time $t_b\propto \thobs^{8/3}$ (where $\thobs$ is the viewing angle), and the pre-break slope can depend on the viewing angle and particular structure model \citep{Rossi:2002aa, Dalal:2002aa, Granot:2002aa, Panaitescu:2003aa, Kumar:2003aa, Granot:2003aa, Salmonson:2003aa, Rossi:2004aa}. The ``Universal Structured Jet'' was a structured jet with isotropic-equivalent energy $E \propto \theta^{-2}$ proposed to explain the diversity of jet-break times as a viewing angle effect  \citep{Lipunov:2001aa, Zhang:2002aa}, although this was ultimately unsuccessful \citep{Nakar:2004aa}. 

The jet break in the structured jet afterglow light curve can exhibit a larger jump between pre- and post-break slopes than in top hat models, and has been invoked to explain some observed jet breaks which do not easily fit the standard closure relations \citep{Panaitescu:2005aa, Panaitescu:2005ab}, although other dynamical and spectral processes can have similar effects \citep{Piro:2005aa, Corsi:2006aa}.  GRB prompt emission viewed significantly off-axis is one of the possible origins for X-ray flashes \citep{Ioka:2001aa, Yamazaki:2002aa, Yamazaki:2003aa, Peng:2005aa, DAlessio:2006aa}.  More recently, small non-zero viewing angles have been measured in a subset of the \swiftXRT{} afterglow sample \citep{Ryan:2015aa, Zhang:2015aa, Troja:2016aa}.  These studies focused on observers positioned within the core opening angle of the jet or relied on closed numerical codes to calculate light curves for misaligned viewers.  

In this paper, we provide a characterization of the afterglows of structured jets at all viewing angles, explicit closure relations, and a description of our  public numerical tool \afterglowpy{}. In Section \ref{sec:motivation} we review the general properties of structured jets and their necessity for GRB afterglow modelling.  In Section \ref{sec:numerical} we develop our theoretical framework and describe its numerical implementation in \afterglowpy{}.  Section \ref{sec:structuredJets} characterizes the general behavior of structured jet afterglow light curves, provides the closure relations and jet break times with explicit viewing angle dependence, and relates structured jets to standard energy injection models.  Section \ref{sec:gw170817} demonstrates an application to \gwbns{}. Section \ref{sec:discussion} gives further discussion and Section \ref{sec:summary} gives the summary.  Detailed derivation of the closure relations is provided in Appendices \ref{app:derive1} and \ref{app:derive2}.

% Explicitly say we were the first to find X-ray, predict steep post-break slope.

%%%%%%%%%%%%%%%%%%%%%%%
%
%  WHY STRUCTURED JETS
%
%%%%%%%%%%%%%%%%%%%%%%%%%

\section{Motivation: Why Structured Jets?}\label{sec:motivation}

The \emph{structured jet} is a GRB jet model where the isotropic-equivalent energy of the blast wave is a function of the angle from the jet axis: $\Eiso = 4\pi dE/d\Omega \equiv E(\theta)$. The particular angular structure of a jet is first imposed by the jet launching mechanism and then modified by the sculpting that occurs as the jet burrows out of the surrounding ejecta debris (as in a binary neutron star merger) or stellar envelope (as in a collapsar).  In the case of neutron star mergers, it is not a given that there is sufficient ejecta in the polar regions to significantly alter the intrinsic jet structure.

\begin{figure}
	\includegraphics[width=\columnwidth]{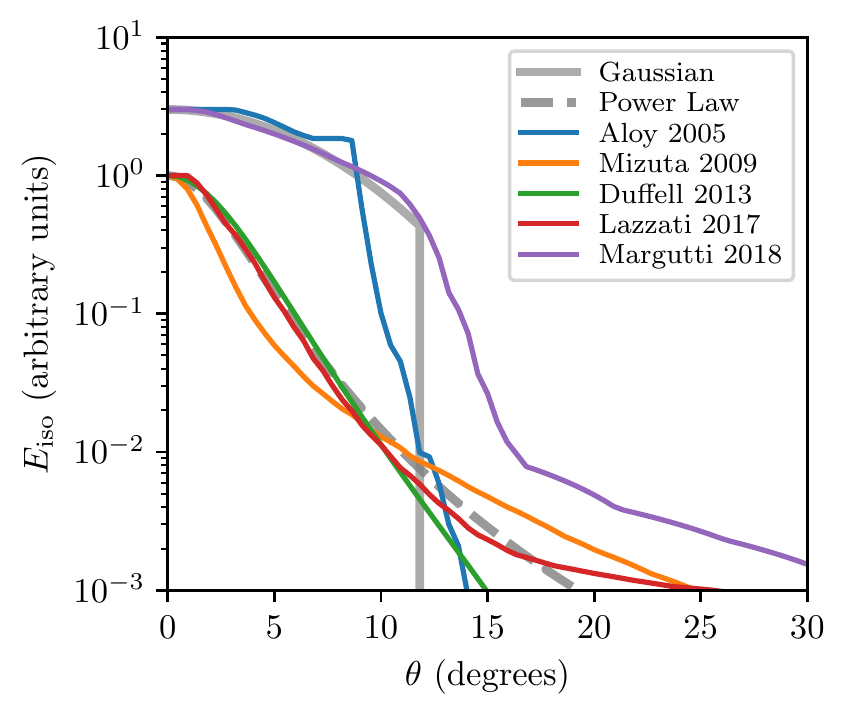}
	\caption{Lateral profiles of isotropic equivalent energy $\Eiso$ as a function of angle from the jet axis $\theta$, individually rescaled to group similar profile shapes.  The thick grey lines show fiducial profiles with simple analytic expressions (Equations \eqref{eq:gauss} and \eqref{eq:pl}) while the thin colored lines show results from numerical simulations and analytic models chosen from the literature.  The blue line is the B1 numerical simulation of \cite{Aloy:2005aa}, orange the HE16N numerical simulation of \cite{Mizuta:2009aa}, green an analytic ``boosted fireball'' model with $\gamma_{\rm{B}}=10$ and $\eta_0=3$ from \cite{Duffell:2013aa}, red the numerical simulation from \cite{Lazzati:2017aa} (L17), and purple the numerical simulation from \cite{Margutti:2018aa} (M18).  The thick solid grey line is a Gaussian profile (Equation \eqref{eq:gauss}) with $E_0 = 3\times 10^{52}$ erg, $\thC = 6^\circ$, and $\thW=12^\circ$ and the thick dashed grey line is a power law profile (Equation \eqref{eq:pl}) with $E_0 = 10^{52}$ erg, $\thC = 2^\circ$, $\thW=20^\circ$, and $b=4.5$.  The Gaussian and power law profiles can emulate the basic properties of the energy profiles found in the literature. \label{fig:jetStruct}}
\end{figure}

Numerical simulations have revealed a variety of jet angular energy distributions, often containing an energetic core with power law tails.  Figure \ref{fig:jetStruct} shows a collection of jet energy distributions from the literature \citep{Aloy:2005aa, Mizuta:2009aa, Duffell:2013aa, Lazzati:2017aa, Margutti:2018aa}.  The \citet{Aloy:2005aa} model (B01 in the paper) was launched as a top hat jet into an accretion torus featuring a narrow underdense funnel region and produced a final energy profile with a relatively sharp edge, although only material with Lorentz factor greater than 100 was included in the $\Eiso$ calculation.  The \citet{Margutti:2018aa} model used a powerful engine with Gaussian injection profile into a standard BNS merger cloud, resulting in a blunt jet with less energetic wings.  The \citet{Mizuta:2009aa} model (their HE16N simulation) is of a collapsar jet, injected with a $5^\circ$ opening angle into a collapsing massive star density profile.  A large degree of interaction with the stellar envelope produces the energetic power law wings.  The \citet{Lazzati:2017aa} model is meant to emulate a BNS merger and injects a top hat jet into a spherically symmetric ejecta wind, also resulting in a large amount of interaction and energetic power law wings.  The \citet{Duffell:2013aa} model is an analytic model of a fireball, spherically symmetric in its rest frame, boosted at a bulk Lorentz factor. It also produces a narrow core with power law wings.

We note that although a variety of jet energy profiles are produced by these works, none except perhaps \citet{Aloy:2005aa} would be considered a ``top hat.''  Simulations of jets launched in realistic environments reliably produce non-trivial lateral energy profiles.  Most GRB jets in nature are likely structured jets.  Given the variation in light curve properties when structured jets are viewed at non-zero inclinations, understanding them is a chief concern for understanding electromagnetic counterparts of gravitational wave sources.

Lacking a well-established physical model of the true $E(\theta)$, in particular its dependence on the parameters of the progenitor system, much of our further discussion considers two simple parameterized models: a Gaussian jet and a power law jet with a smooth core.  

\begin{align}
	E(\theta) &= E_0 \exp\left(-\frac{\theta^2}{2\thC^2}\right)  && \text{Gaussian} \label{eq:gauss}\\
	E(\theta) &= E_0 \left(1 + \frac{\theta^2}{b\thC^2} \right)^{-b/2}  && \text{power law} \label{eq:pl}
\end{align}

Each fiducial model is parameterized by a normalization $E_0$, a width $\thC$, and a truncation angle $\thW$ outside of which the energy is initially zero.  The power law model also retains a power law index $b$.     Our power law profile includes a factor of $b^{-1}$ in the base which does not usually appear in the literature \citep{Granot:2003aa, Hotokezaka:2018aa}.  This serves to normalize the value of $\thC$ and make it comparable between power laws of different $b$ and Gaussian jets.  Both structure profiles obey
\begin{equation}
	\left . \frac{d^2}{d\theta^2} \log E \right \rvert_{\theta=0} = -\frac{1}{\thC^2}\ , \label{eq:thCdef}
\end{equation} 
which we take as a generic definition for $\thC$.

Figure \ref{fig:jetStruct} also includes an example of each model together with the jet energy profiles drawn from the literature.  The power law model emulates those jets subject to a large degree of interaction with energetic wings.  The Gaussian model emulates jets which experienced less (or more focused) interaction and retain sharper sides.  As $b$ increases the power law profile gets steeper and the wings become less energetic, until in the limit $b\to \infty$ the Gaussian profile is recovered.

A generic energy profile $E(\theta)$, for instance from a numerical hydrodynamics simulation, may have several more parameters, but one can still associate with it an on-axis energy $E_0 = E(0)$ and an effective core width $\thC \sim |E''(0)/E_0|^{-1/2}$ which should fully specify the leading order near on-axis behaviour.  

The Lorentz factor profile $\gamma(\theta)$ of a jet may also obtain angular structure through its launching and evolution.  In this work we focus on the consequences of non-trivial $E(\theta)$, as the bulk of successful observations are expected to occur when the jet is in the deceleration regime and the jet is locked to $\gamma(\theta) \propto E(\theta)^{1/2}$.  We leave the specific consequences of differing $\gamma(\theta)$ profiles, which may affect emission at early times, to future work.

%%%%%%%%%%%%%%%%%%%%%%%
%
%  METHODS
%
%%%%%%%%%%%%%%%%%%%%%%%%%

\section{Methods And The \afterglowpy{} Package}\label{sec:numerical}

To compute the light curves of structured jet afterglows we constructed numerical and analytic models utilizing the single shell approximation of \cite{van-Eerten:2010aa, van-Eerten:2018ab}.  Iterations of this model have been applied in the \gwbns{} X-ray discovery paper \citep{Troja:2017aa}, follow-up studies of \gwbns{} \citep{Troja:2018aa, Piro:2019aa, Troja:2019ab}, and in examinations of archival kilonova candidates \citep{Troja:2018ab, Troja:2019aa}. This approach integrates over the massive ejecta, contact discontinuity, and forward shock complex, treating it as a single fluid element with uniform radial structure. We utilize a trans-relativistic equation of state which smoothly interpolates between the ultra-relativistic and non-relativistic limits \citep{van-Eerten:2013ab, Nava:2013aa} and include an approximate prescription for jet spreading.  This approach, with a simplified equation of state and jet spreading model, has been used successfully to model the synthetic light curves of top hat jets from multidimensional numerical relativistic hydrodynamics simulations \citep{van-Eerten:2010aa}.

In the ultra-relativistic limit the single shell approximation provides useful scaling relations to compute the structured jet closure relations presented in Section \ref{sec:structuredJets}.  The full trans-relativistic numerical model is publicly available as the \afterglowpy{} Python package, described in more detail in Section \ref{subsec:afterglowpy}.

We utilize a standard spherical coordinate system $(r, \theta, \phi)$ with origin at the GRB central engine and polar axis aligned with the jet axis.  The blast wave forward shock has a radial position $R(t, \theta)$, where $t$ is the time measured in the burster frame. The observer is located in a direction $\hat{\bf n}$ which makes an angle $\thobs$ with the $z$-axis, $\hat{\bf n} \cdot \hat{\bf z} = \cos \thobs$, and is oriented along the $x$-axis, $\phobs = 0$.  A particular point on the blast wave $\hat{\bf r} = (\theta, \phi)$ makes an angle $\psi$ (with cosine $\mu$) with the viewer direction, $\mu = \cos \psi = \hat{\bf n} \cdot \hat{\bf r}$.

\subsection{The Single Shell Approximation}\label{subsec:algo}

The observed flux $F_\nu(\tobs, \nuobs)$ at observer time $\tobs$ and frequency $\nuobs$ is calculated via:
\begin{equation}
	F_\nu(\tobs, \nuobs) = \frac{1+z}{4\pi \dL^2} \int \! \dd \Omega\  \dd r\ r^2 \delta^2\ \epsilon'_{\nu'} \ , \label{eq:flux}
\end{equation}
where $z$ is the redshift of the source, $\dL$ is the luminosity distance, $\delta$ the doppler factor of the emitting fluid with respect to the observer, and $\epsilon'_{\nu'}$ the fluid rest-frame emissivity.

To accommodate an initial structure profile $E(\theta)$ we consider the integrand of Equation \eqref{eq:flux} as a function of the polar angle $\theta$.  We assume each constant-$\theta$ annulus evolves independently, as an equivalent top hat of initial width $\theta_j = \theta$.  This is a very good approximation when transverse velocities are low: when the blast wave is ultra-relativistic and has not begun to spread and when the blast wave is non-relativistic and the spreading has ceased \citep{van-Eerten:2010aa}.  Once jet spreading begins in earnest the errors are larger, and this approach can be best viewed as an interpolation between the correct ultra-relativistic and non-relativistic limits \citep{van-Eerten:2010aa}.
 
To compute the light curve of a top-hat jet of initial width $\theta_0$ we first must calculate the time evolution of the blast wave. In the single-shell approximation we treat the ejecta mass, contact discontinuity, and forward shock as a single unit propagating through a cold ambient medium with constant rest-mass density $\rho_0 = \Mp n_0$.  We utilize the ``TM'' trans-relativistic equation of state to describe the fluid consistently throughout its evolution \citep{Mignone:2005aa}.  The forward shock is at radius $R$ from the explosion, and the fluid behind the shock has dimensionless four-velocity $u$ and Lorentz factor $\gamma$.  The shock jump conditions can be used to determine the shock speed:
\begin{equation}
	\dot{R} = \frac{4 u \gamma}{4 u^2 +3}c\ . \label{eq:Rdot}
\end{equation}
Here the time derivative $\dot{R}$ is taken with respect to elapsed time in the bursters' frame $t$.  

 The evolution of the four velocity is determined through conservation of energy.  The total energy in the trans-relativistic single shell approximation is:
\begin{align}
	E &= (\gamma - 1)\Mej c^2 + \frac{4\pi}{9} \rho_0 c^2 R^3 (4 u^2 + 3) \beta^2 f_\Omega \ , \label{eq:E} \\
	f_\Omega &= 2 \sin^2\left( \theta_j / 2 \right)\ . \label{eq:fOm}
\end{align}

In Equation \eqref{eq:E} the first term is the kinetic energy of the ejected mass $\Mej$, assumed to have already accelerated and adiabatically cooled to its coasting velocity. The second term is the kinetic and thermal energy of the shocked ISM with three-velocity $\beta = u / \gamma$.  Equation \eqref{eq:fOm} describes the fractional solid angle of the jet in terms of the time dependent opening angle $\theta_j(t)$. 

Pressure gradients along the blast wave surface can drive lateral spreading of the jet, causing $\theta_j(t)$ (and $f_\Omega$) to increase with time  from its initial value $\theta_j(0) = \theta_0$ \citep{Rhoads:1999aa}.  Initially spreading is negligible as each fluid element of the highly relativistic jet is only in causal contact with nearby regions of similar pressure.  Spreading begins once sound waves have had sufficient time to travel across the surface of the jet and communicate the presence of pressure gradients, launching rarefaction waves which drive the jet to spread at its local sound speed \citep{Rhoads:1999aa, van-Eerten:2012aa, Duffell:2018aa}.  Spreading continues until the blast wave becomes spherical and $f_\Omega = 1$.

Numerous semi-analytic prescriptions for jet spreading exist in the literature, for example \citet{Rhoads:1999aa} (R99), \citet{Granot:2012aa} (GP12), and \citet{Duffell:2018aa} (DL18).  To remain consistent with the presence of angular structure in the jet and ensure spreading occurs at sonic speeds in our adopted equation of state it is necessary to construct our own jet spreading prescription.  

The expression for the jet energy, Equation \eqref{eq:E}, implicitly assumes the ``conical'' spreading model, where at a time $t$ all material $r < R(t)$ and $\theta < \theta_j(t)$ is incorporated in the jet.  Comparisons to numerical simulation have found this to be more accurate than the ``trumpet'' model which only incorporates material that was within the jet cone at each radius (DL18).

A wave traveling at the sound speed $c_s$ in the fluid rest frame will maintain a velocity $\beta_{\bot} c$ laterally along the shock front:
\begin{equation}
	\beta_\bot = \gamma \sqrt{ (1-\beta \dot{R}/c)c_s^2 - (\dot{R}/c - \beta)^2} \ .
\end{equation}
Using the shock-jump conditions and TM equation of state this reduces to:
\begin{equation}
	\beta_\bot = \sqrt{\frac{2u^2+3}{4u^2+3}} \frac{\dot{R}}{2\gamma c}\ . \label{eq:spreadVel}
\end{equation}
In the ultra-relativistic limit $\beta_\bot = \sqrt{3/8} c_s/(c\gamma) = 1/(\sqrt{8} \gamma)$.  In this same limit, a signal launched at $t=0$ traveling at $\beta_\bot$ along the shock front will cover an angular distance  $\Delta \theta = 1/(3\sqrt{2}\gamma)$ \citep{van-Eerten:2012aa}.  Taking $\thC$, defined by Equation \eqref{eq:thCdef}, as the angular scale for order unity changes in the jet pressure, we can approximate the onset of spreading as when $u = 1/(3\sqrt{2}\thC)$.  Further taking $\beta_\bot$ as the expansion velocity once spreading begins, we can write an evolution equation for $\theta_j$:
\begin{equation}
	\dot{\theta}_j = \begin{cases}
					0 & \text{if } u > 1/(3\sqrt{2} \thC)\\
					\beta_\bot c / R &\text{otherwise}
				\end{cases}\ . \label{eq:thetadot}
\end{equation}
The spreading prescription in Equation \eqref{eq:thetadot} is very similar to existing models in the literature.  Like R99, GP12 ($a=0$ case), and DL18 in the ultra-relativistic limit we have $\dot{\theta}_j \propto 1/(\gamma t)$.  Like DL18 we freeze spreading completely while $u \thC$ exceeds a threshold value.  In fact, while $u \gg 1$ our prescription is equivalent to the DL18 model with parameters $P_k = Q_k = 3\sqrt{2}$, a slight modification to their numerically fit values $P_k = 4.0$ and $Q_k = 2.5$.  Afterglow light curves computed using Equation \eqref{eq:thetadot} were found to agree better with established simulation based models (see Section \ref{subsec:boxfitcomp}) than those computed with the DL18 parameters.  This jet spreading prescription differs from that used in \citet{Troja:2018aa, Piro:2019aa, Troja:2019ab} which used a simple $u < 1$ criterion to begin jet spreading and set $\beta_\bot = c_s / (c \gamma)$, following \citet{van-Eerten:2010aa}.

Given an expression for the total blast wave energy $E$ in terms of $t$, $R$ and $u$, Equations \eqref{eq:Rdot}, \eqref{eq:E}, and \eqref{eq:thetadot} are sufficient to write a system of ordinary differential equations (ODEs) in $R(t)$, $u(t)$, and $\theta_j(t)$. A standard adiabatic blast wave maintains constant $E$, while energy injection and refreshed shock models may add dependence on $t$ or $u$. The general form of the shock evolution equations are then:
\begin{align}
	\dot{R} &= \frac{4 u \gamma}{4 u^2 +3}c  \label{eq:ODE1}\\
	\dot{\theta}_j &= \frac{1}{2\gamma}\sqrt{\frac{2u^2+3}{4u^2+3}} \frac{\dot{R}}{R}  \label{eq:ODE2} \\
	\dot{u} &= -\frac{E_{sw}(4u^2+3)\beta^2 (3\dot{R}/R + \cot (\theta_j/2) \dot{\theta}_j) - \partial_t E}{2 E_{sw} u (4u^4 + 8u^2 + 3)\gamma^{-4} + \beta \Mej c^2 - \partial_u E}  \label{eq:ODE3} \\
	E_{sw} &\equiv \frac{4\pi}{9} \rho_0 c^2 R^3 f_\Omega\ . \nonumber
\end{align}
Equations \eqref{eq:ODE1}-\eqref{eq:ODE3} define a three dimensional system of ODEs in the variables $(R, u, \theta_j)$ which may be solved numerically or, in certain limits, analytically from appropriate initial conditions.  Once the shock evolution ($R(t)$, $u(t)$, $\theta_j(t)$) is known, the flux is given by the integral:
\begin{align}
	F_\nu(\tobs, \nuobs) &= \frac{1+z}{4\pi \dL^2} \int \! \dd \Omega\ R^2\ \Delta R\  \delta^2\ \epsilon'_{\nu'} \ , \label{eq:flux2}
\end{align}
where $\Delta R$ is the effective shock width contributing to the emission \citep{van-Eerten:2010aa}:
\begin{equation}
	\Delta R = \frac{1}{1-\mu \dot{R}} \frac{R}{12\gamma^2} \ . \label{eq:dr}
\end{equation}
 The integrand is evaluated at a constant observer time $\tobs$ and observer frequency $\nuobs$, related to $t$ and $\nu'$ by:
\begin{align}
	\tobs &= (1+z) \left(t - \mu(\theta, \phi) R(t)/c\right)\ , \label{eq:tobs} \\
	\nuobs &= (1+z)^{-1} \delta \nu' , \label{eq:nuobs}
\end{align}

The rest-frame synchrotron emissivity $\epsilon'_{\nu'}$ may be calculated with varying levels of sophistication.  We use the standard broken power law formalism with characteristic frequencies $\nu_m$ and $\nu_c$, with the cooling frequency $\nu_c$ calculated via the global cooling approximation \citep{Granot:2002aa, van-Eerten:2010aa}.  The fluid behind the shock has rest-frame number density $n' = 4n_0\gamma$, thermal energy $e'_{th} = (\gamma-1)m_p n' c^2$, and magnetic field strength $B = \sqrt{8\pi e'_{th} \epsB}$, where $\epsB$ is the fraction of thermal energy in the magnetic field. We assume a fraction $\xiN$ of the electron population is shock-accelerated electrons with a fraction $\epse$ of the thermal energy and have a Lorentz factor distribution $N(\gamma_e) \propto \gamma_e^{-p}$ with index $p>2$.

Each characteristic frequency $\nu_i$ has a corresponding Lorentz factor $\gamma_i$:
\begin{equation}
	\nu_i = \frac{3e B}{4 \pi m_e c}  \gamma_i^2 \ ,
\end{equation}
where $e$ is the elementary charge.  These characteristic Lorentz factors are:
\begin{align}
	\gamma_m &= \frac{2-p}{1-p} \frac{\epse e'_{th}}{\xiN n' m_e c^2}\ , \\
	\gamma_c &= \frac{6\pi m_e \gamma c}{\sigma_T B^2 t}\ .
\end{align}
The synchrotron spectrum has peak emissivity $\epsilon_P$:
\begin{equation}
	\epsilon_P = \frac{p-1}{2}\frac{\sqrt{3} e^3 \xiN n' B}{m_e c^2}\ .
\end{equation}
Finally, the rest-frame emissivity $\epsilon'_{\nu'}$ is given as:
\begin{equation}
	\epsilon'_{\nu'} = \epsilon_P \times \begin{cases}
					(\nu' / \nu_m)^{1/3} & \nu' < \nu_m < \nu_c \\
					(\nu' / \nu_m)^{-(p-1)/2} & \nu_m < \nu' < \nu_c \\
					(\nu_c / \nu_m)^{-(p-1)/2}(\nu' / \nu_c)^{-p/2} & \nu_m < \nu_c < \nu' \\
					(\nu' / \nu_c)^{1/3} & \nu' < \nu_c < \nu_m \\
					(\nu' / \nu_c)^{-1/2} & \nu_c < \nu' < \nu_m \\
					(\nu_m / \nu_c)^{-1/2}(\nu' / \nu_m)^{-p/2} & \nu_c < \nu_m < \nu'
				\end{cases} \label{eq:syncSpec}
\end{equation}

\subsection{\afterglowpy{}} \label{subsec:afterglowpy}

We have constructed the \afterglowpy{} Python package to implement the numerical computation of light curves according to Section \ref{subsec:algo} and provide it to the community.  The integration routine itself is written in C, wrapped as an extension for Python, and has been optimized to be used in intensive data analysis routines such as Markov Chain Monte Carlo which can require many thousands or millions of evaluations.  

\afterglowpy{} uses the standard fourth-order Runge-Kutta algorithm to numerically solve the $(R(t), u(t), \theta_j(t))$ system of ordinary differential equations (Equations \eqref{eq:ODE1}-\eqref{eq:ODE3}) on a fixed logarithmically spaced grid of $t$ \citep{Press:2007aa}.  The endpoints of the $t$-grid are chosen to bracket the burster frame times required to calculate the requested $\tobs$.  Initial conditions for $R$, $u$, and $\theta$ are typically those of a decelerating ultra-relativistic blast wave.  The user can set the density of the $t$-grid with the {\tt tRes} parameter: the number of grid points per decade of $t$.  The default value for {\tt tRes} is 1000, which is sufficiently dense that the blast wave ODE evolution is not the dominant error source but not so dense as to adversely impact performance.

Each top-hat jet component is integrated in $\theta$ and $\phi$ using an adaptive Romberg scheme with a fixed relative tolerance of $10^{-6}$ and an adaptive absolute tolerance.  When summing over top-hat components of a structured jet, the innermost (core) component is calculated first.  As the calculation proceeds, the current running sum of the flux is used to set absolute tolerance for the next component.  This minimizes the computations performed on dim sectors of the jet far from the line of sight.

Each evaluation of the integrand requires a binary search to determine the burster time $t$ at which to evaluate $R$, $u$, and $\theta_j$.  Fluid quantities are then calculated using the shock-jump conditions and the synchrotron emissivity is evaluated using the standard external shock formulae (Equation \eqref{eq:syncSpec}, see also \citet{Granot:2002aa,van-Eerten:2010aa}).  

The numerical accuracy of a top-hat light curve with this integration scheme is typically better than $10^{-4}$.  The structured jet calculation splits the integration domain into {\tt latRes} disjoint annuli, each evolved as an independent top-hat and their emission summed.  By default {\tt latRes} is chosen such that there will be 5 zones per $\thC$-sized interval of $\theta$.   The choice of  {\tt latRes} has the largest impact on the code performance and accuracy.  We find the choice $\mathtt{latRes} = 5$ gives sufficiently quick performance at acceptable errors, typically on the order of $10^{-2}$.

\afterglowpy{} is available on PyPI and may be installed with {\tt pip} as {\tt pip install afterglowpy}.  The source code is open and available at \url{https://github.com/geoffryan/afterglowpy}.

\subsection{Comparison To \boxfit{}} \label{subsec:boxfitcomp}

\afterglowpy{}, in utilizing the semi-analytic methods of Section \ref{subsec:algo}, trades some amount of physical accuracy for great flexibility.  We gauge this trade-off by comparing to the \boxfit{} code, a standard tool which calculates high fidelity afterglow light curves based on numerical simulations \citep{van-Eerten:2012ab}.  

\boxfit{} uses two-dimensional relativistic hydrodynamic simulations to fully capture the non-linear hydrodynamics of a decelerating blast wave and a ray tracing radiative transfer module to compute observed synchrotron light curves.  The simulations begin from a top hat jet initial condition with a Blandford-Mckee radial profile \citep{Blandford:1976aa}, are evolved adiabatically with the trans-relativistic TM equation of state, and fully capture the lateral spreading of the jet \citep{van-Eerten:2012ab}.  Comparison to \boxfit{}, with its much more accurate hydrodynamic evolution, tests the fidelity of the assumptions and approximations of the single shell approach used by \afterglowpy{} and is a very useful calibration point.  

%These errors quoted are numerical errors associated with the schemes used to solve Equations \ref{eq:Rdot}, \ref{eq:E}, and \ref{eq:thetadot} and perform the integral in Equation \ref{eq:flux2}.  That is, they are the errors relative to solving these equations perfectly.  The equations themselves are, however, an approximation themselves to the full physics of an afterglow light curve.  Comparison and calibration against analytic calculations and hydrodynamic simulations is necessary to ensure physically realistic results.  
%
\begin{figure*}
	\includegraphics[width=\textwidth]{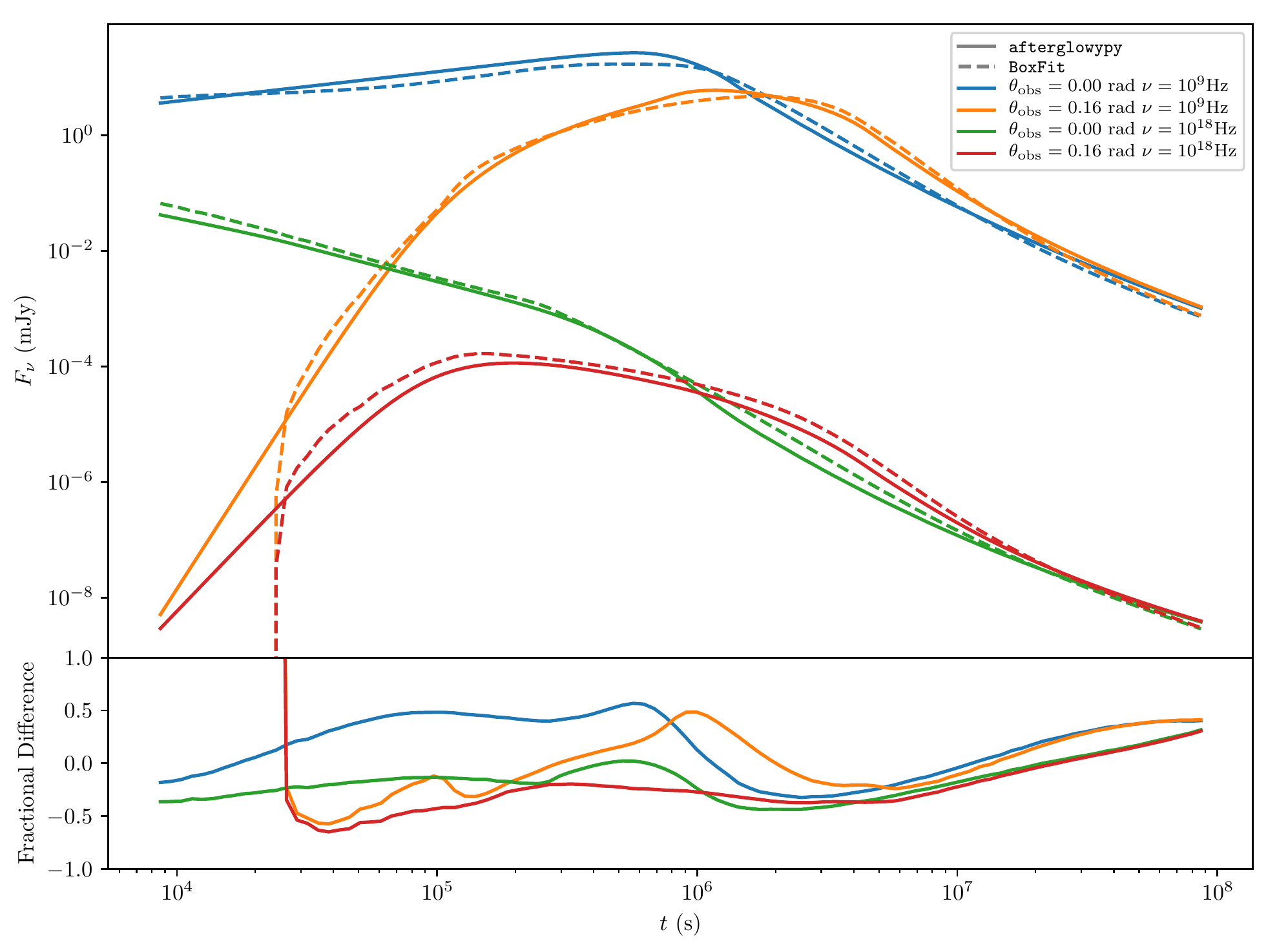}
	\caption{Top Panel: Comparison between top hat jet light curves from \afterglowpy{} (solid lines) and \boxfit{} (dashed lines). Bottom Panel: Fractional difference between \afterglowpy{} and \boxfit{} light curves.  Four representative light curves are shown: radio aligned ($\thobs=0$, $\nu=10^9$Hz, blue), radio misaligned ($\thobs=0.16$rad, $\nu=10^9$Hz, orange), X-ray aligned ($\thobs=0$, $\nu=10^{18}$Hz, green), and X-ray misaligned ($\thobs=0.16$rad, $\nu=10^{18}$Hz, red).  Remaining parameters are shared: $\thC=0.1$rad, $\Eiso=10^{52}$erg, $n_0 = 10^{-3}$cm$^{-3}$, $p=2.2$, $\epse=10^{-1}$, $\epsB=10^{-2}$, $\dL=3.09 \times 10^{26}$cm, $z=0.028$. \label{fig:boxfitComp}}
\end{figure*}

Figure \ref{fig:boxfitComp} shows a comparison between top-hat light curves calculated with \afterglowpy{} and \boxfit{}.   Figure \ref{fig:boxfitComp} shows the light curve of a $\thC=0.1$ rad top-hat jet at radio and X-ray frequencies ($10^9$ Hz and $10^{18}$ Hz, respectively) with aligned ($\thobs = 0$) and misaligned ($\thobs=0.16$ rad) viewing angles.  The light curves begin in the ultra relativistic Blandford-McKee phase and continue to the Newtonian Sedov phase.

The overall agreement is good: \afterglowpy{} captures the salient features of both aligned and misaligned light curves including the jet break, spectral shape, transition to the Sedov phase, and the early steep rise for misaligned viewing.  
At early times, \boxfit{} lacks the simulation coverage to produce accurate fluxes, leading to a large discrepancy.  Once proper coverage is attained \boxfit{} and \afterglowpy{} show less than 50\% relative discrepancy, apart from the on-axis radio light curve which briefly has a 60\% discrepancy at the time of the jet break ($t_b \approx 7 \times 10^5$ s for the aligned light curves, $t_b \approx 3\times 10^6$ s for the misaligned).  After the jet break the \afterglowpy{} light curves recover the \boxfit{} slope, and asymptote to similar light curves in the Newtonian regime with a small constant offset from \boxfit{}.

As expected, \afterglowpy{} models the early and late afterglow well, but is somewhat less accurate in the intermediate phase during the onset of jet spreading. This phase is precisely where the non-linear hydrodynamics of the blast wave are most important, and hardest to model with simple semi-analytic approximations.  Any work utilizing \afterglowpy{} should be aware of this fact, and be careful when treating data from this regime.

It bears pointing out that some of the hard discrepancy between the codes is due to the test itself: the sharp-edged nature of the top-hat jet exacerbates discrepancies in how these edges are treated.  An angularly structured jet with non-trivial $E(\theta)$ may smooth over the differences in approach between the codes: we may expect \afterglowpy{} to be more accurate for structured jets than top-hats.  \afterglowpy{} is by no means a replacement for full numerical simulations, but is a useful and flexible tool which captures much of the important physics of GRB afterglows at a very small fraction of the cost of a relativistic numerical hydrodynamic simulation.

%%%%%%%%%%%%%%%%%%
%
% Afterglows Section
%
%%%%%%%%%%%%%%%%%%%%

\section{Structured Jet Light Curves}\label{sec:structuredJets}

Afterglow light curves, even when viewed off axis, can typically be described as a piecewise set of power law segments $F_\nu \propto \tobs^\alpha \nuobs^\beta$, each denoting particular phases of evolution of the blast wave and its emitting region.  Particular phases have been identified empirically \citep{Nousek:2006aa} and theoretically \citep{Zhang:2006aa}.   Within a particular phase, the slopes $\alpha$ and $\beta$ are often related through a \emph{closure relation}: an equality determining $\alpha$ from $\beta$ or vice versa.  Phases change at \emph{break times} when the temporal slope $\alpha$ (and potentially the spectral slope $\beta$) transition rapidly from the previous phase to the next.

To fully characterize the light curves of structured jets we use \afterglowpy{} to construct light curves exploring dependence on jet structure model, viewing angle $\thobs$, opening angle $\thC$, and synchrotron regime.  We identify the relevant phases in the afterglow and present new closure relations and jet break time scalings for the \emph{structured} phase of evolution: when the patch of the blast wave dominating the emission is sliding from the line of sight towards the jet core.

Previous work has demonstrated the light curves of structured jets show two modes of behaviour, depending on whether the observer is \emph{aligned} ($\thobs < \thC$) or \emph{misaligned} ($\thobs > \thC$) \citep{Rossi:2002aa, Granot:2002ab, Panaitescu:2003aa, van-Eerten:2010aa}.
In the \emph{aligned} case, the light curves follow the standard on-axis top hat behaviour modified slightly for non-zero viewing angle.  There is little to distinguish between different structure $E(\theta)$, and the light curve is well-approximated by a broken power law with characteristic break times \citep{Granot:2002aa}.  The \emph{misaligned} light curve structure is more complicated.  It also presents as a broken power law, but with closure relations explicitly dependent on viewing angle and jet angular structure.  

Due to relativistic beaming the flux at any given time is dominated by a small patch of blast wave surface.  The angular coordinates of this patch are $\theta = \theta_*$ and $\phi=0$, it has angular size $\Delta \Omega$, Lorentz factor $\gamma_*$, and emits a specific intensity $I_\nu^*$. We can write the received flux as:
\begin{equation}
	F_\nu = I_\nu^*\ \Delta \Omega\ .
\end{equation}

The behaviour of $\theta_*$ and $\Delta \Omega$ control the overall light curve evolution.  We encode the dependence of $\Delta \Omega$ on the blast wave Lorentz factor as:
\begin{equation}
	\Delta \Omega \propto \gamma_*^{-\som} \ .
\end{equation}
The parameter $\som$ controls the growth rate of the visible patch, and is typically between 0 and 2. 

\subsection{Phases Of Evolution} \label{subsec:phases}

\begin{figure*}
	\includegraphics[width=\textwidth]{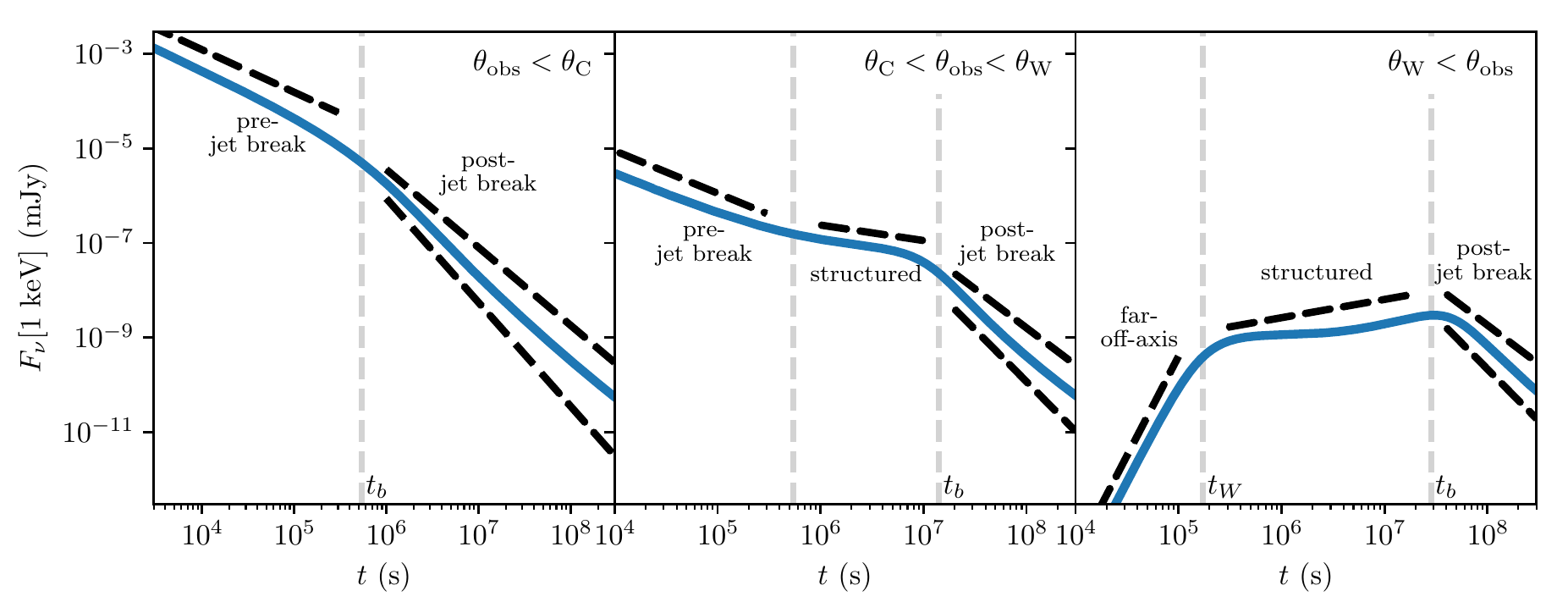}
	\caption{Illustration of possible synchrotron light curves from the forward shock of a structured jet.  The specific order of light curve phases depends on where the viewer sits, $\thobs$, relative to the jet half opening angle $\thC$ and the outer truncation angle $\thW$. A \emph{coasting} phase, ending at $\tdec$, is assumed to have already completed.  Each panel shows a fiducial light curve (solid blue lines), power law segments $t^\alpha$ with slopes from the closure relations in Table \ref{tab:slopes} (black dashed lines), and representative break times between each phase of evolution (grey dashed vertical lines). The two dashed lines in the ``post-jet break'' section bracket the possible light curve behaviour.  The viewing angle $\thobs = 0$ rad, $0.2$ rad, and $0.4$ rad in the left, center, and right panels respectively.  Remaining parameters are shared: $\thC=0.05$ rad, $\thW=0.3$ rad, $\Eiso=10^{53}$ erg, $n_0 = 10^{-3}$ cm$^{-3}$, $p=2.2$, $\epse=10^{-1}$, $\epsB=10^{-3}$, $\dL=10^{28}$ cm, $z=0.5454$. 
	\label{fig:phaseSketch}}
\end{figure*}

The afterglow of a structured jet evolves through several phases, depending on the observer orientation, jet structure, and Lorentz factor of the blast wave.  We summarize each phase below in roughly temporal order.  Table \ref{tab:slopes} gives the \emph{closure relations}, the temporal and spectral power law slopes of the observed flux, for each of the afterglow phases in the synchrotron spectral regimes D-H (as defined in \citet{Granot:2002aa}). Figure \ref{fig:phaseSketch} sketches the three possible ways these phases may be combined in a full afterglow, depending on the viewing angle.

\begin{deluxetable*}{CCCCCCCCC}
	\tablecaption{Structured jet closure relations: $F_\nu \propto \tobs^{\alpha} \nuobs^{\beta}$. The ``far-off-axis'' phase ($\alpha_{\mathrm{far-off-axis}}$) occurs at early times and only if there is no relativistic material directly in the observer's line of sight ($\thobs > \thW$). The ``generic'' slope ($\alpha_{\mathrm{generic}}$) applies whenever the jet is relativistic, non-spreading, and the observer is within the beaming cone of some part of the jet ($|\thobs - \theta_*| < \gamma^{-1}(\theta_*)$). The $g$ parameter is positive and depends on the viewing angle $\thobs$ and the specific structure profile $E(\theta)$.  See Equations \eqref{eq:gdef} and \eqref{eq:geff} for the definition and effective values of $g$.  For a top hat jet $g=0$ and for a Gaussian jet $g \approx \thobs^2/(4\thC^2)$. The $\som$ parameter encodes how the effective angular size of the jet is changing with Lorentz factor: $\Delta \Omega \propto \gamma^{-\som}$.   The ``pre-jet break'' phase ($\alpha_{\mathrm{pre}}$) is seen by aligned viewers ($\thobs \lesssim \thC$) and by early misaligned viewers if there is high Lorentz factor material in their line of sight.  The ``structured'' phase ($\alpha_{\mathrm{struct}}$) is seen by misaligned viewers ($\thobs > \thC$), occurs before the jet break, and corresponds to $\som \approx 1$.   The ``post-jet break'' slopes ($\alpha_{\mathrm{post, 1}}$ and $\alpha_{\mathrm{post, 2}}$) bracket the possible post-jet break behaviour.  The first is for a non-spreading jet while the second corresponds to a jet undergoing exponential spreading and was calculated \citet{Sari:1999aa}.\label{tab:slopes}}
	\tablehead{\colhead{Regime}& \colhead{Label} & \colhead{$\alpha_{\mathrm{far-off-axis}}$} & \colhead{$\alpha_{\mathrm{generic}}$} & \colhead{$\alpha_{\mathrm{pre}}$} & \colhead{$\alpha_{\mathrm{struct}}$}& \colhead{$\alpha_{\mathrm{post, 1}}$} & \colhead{$\alpha_{\mathrm{post, 2}}$} & \colhead{$\beta$}}
	\startdata
	& & & & \som = 2 & \som=1 & \som=0 &\\
	& & & &  g = 0 & & g=0 & \\[0cm]
	\nu<\nu_m<\nu_c     & D	& 7 		& \dfrac{-2 + 3\som + 3g}{8+g} 		&1/2	& \dfrac{1 + 3g}{8+g} 		& -1/4	& -1/3 & 1/3 \\[0.4cm]
	\nu < \nu_c < \nu_m & E	& 17/3 	& \dfrac{-14/3 + 3\som + 11g/3}{8+g}  &1/6 & \dfrac{-5/3 + 11g/3}{8+g} 	& -7/12	& -1    & 1/3 \\[0.4cm]
	\nu_c < \nu < \nu_m & F	& 13/2 	& \dfrac{-8 + 3\som + 2g}{8+g} 		& -1/4 & \dfrac{-5 + 2g}{8+g} 		& -1		& -1	  & -1/2 \\[0.4cm]
	\nu_m<\nu<\nu_c     & G 	&\dfrac{15 - 3p}{2} & \dfrac{-6p + 3\som + 3g}{8+g} & -\dfrac{3(p-1)}{4}	& \dfrac{3 - 6p +3g}{8+g} 	 & -\dfrac{3p}{4}	& -p	  & \dfrac{1-p}{2} \\[0.4cm]
	\nu_m, \nu_c < \nu & H 	& \dfrac{16 - 3p}{2} 	& \dfrac{-6p - 2 + 3\som + 2g}{8+g} 	& -\dfrac{3p-2}{4} & \dfrac{1-6p  +2g}{8+g} 	& -\dfrac{3p+1}{4} & -p & -\dfrac{p}{2} \\[0.4cm]
	\enddata
\end{deluxetable*}

\emph{Coasting} --- Before the blast wave begins to decelerate it coasts at constant Lorentz factor. The flux depends only on the total volume of the emitting region and the evolution of the cooling frequency $\nu_c$. 

\emph{Far Off Axis} --- If the viewer lies outside the truncation angle, $\thobs > \thW$, then at early times there is no material in the line of sight, the entire blast wave surface is off-axis, and all emission is beaming suppressed.  This greatly reduces the early flux and leads to a dim but steeply rising light curve as the blast wave decelerates.  Emission is dominated by material on the edge nearest the observer, $\theta_* = \thW$, and the angular size of the visible patch is constant in time, $\som = 0$.  This phase ends at $\tW$ when $\gamma^{-1}(\thW) \sim \thobs-\thW$.  If $\thobs < \thW$ this phase is entirely absent.  Table \ref{tab:slopes} gives the temporal power law slope of the light curve in this phase.  These slopes are derived in Appendix \ref{app:derive1}.

\emph{Pre-jet break} --- The standard on-axis early afterglow phase.  The flux is dominated by a small patch of relativistic material directly in the line of sight: $\theta_* = \thobs$.  The material is relativistic, $\gamma_* \gg 1$, and the angular size is controlled by doppler beaming: $\Delta \Omega \sim \gamma_*^{-2}$ and $\som = 2$.  The patch is sufficiently small (or sufficiently on-axis) that $E(\theta)$ is near-uniform over the visible patch.  Closure relations and scalings of the standard on-axis top hat jet apply, e.g. \citet{Granot:2002aa}.  This phase occurs for all aligned observers, $\thobs < \thC$.  It may also occur at early times for misaligned viewers with material in the line of sight, $\thC < \thobs < \thW$, if the Lorentz factor in the line of sight $\gamma_*$ is sufficiently high.

\emph{Structured} --- If the viewer lies outside the core of the blast wave, $\thobs > \thC$, non-trivial angular structure can have a significant effect on the light curve.  Material from the energetic regions progressively closer to the jet axis continuously comes into view, becoming the new dominant source of emission, and moving the centroid with time: $\theta_* \to 0$ as $\tobs$ increases.  The centroid obeys $\gamma^{-1}(\theta_*) \sim |\thobs-\theta_*|$.  The angular size of the visible patch in the azimuthal direction is still controlled by beaming, but in the polar direction it is controlled by the steep gradient in $E(\theta)$, resulting in $\som \approx 1$.  The temporal evolution of the light curve depends on both the specific angular structure $E(\theta)$ and the viewing angle $\thobs$. The specific closure relations are given in Table \ref{tab:slopes} and discussed in Section \ref{subsec:structure}.

\emph{Post-jet break} --- The jet break is an achromatic break in the light curve caused when the angular size of the visible patch ceases to change: $\som = 0$.  At the same time, jet spreading begins in earnest causing the jet to decelerate quicker \citep{Rhoads:1999aa}.  Both effects cause an achromatic steepening of the light curve.  Table \ref{tab:slopes} gives an over-estimate of the post-jet break slope, including the $\som=0$ effect but neglecting jet spreading.

Figure \ref{fig:phaseSketch} shows the three possible orderings of these phases, depending on the viewing angle and particular jet structure.  The ``standard afterglow'' occurs for aligned observers, $\thobs < \thC$, and follows \emph{coast} $\to$ \emph{pre-jet break} $\to$ \emph{post-jet break}.  Misaligned viewers inside the truncation angle, $\thC < \thobs < \thW$, will see \emph{coast} $\to$ \emph{pre-jet break} $\to$ \emph{structured} $\to$ \emph{post-jet break}.  Misaligned viewers outside the truncation angle, $\thobs > \thC, \thW$, will see \emph{coast} $\to$ \emph{far off-axis} $\to$ \emph{structured} $\to$ \emph{post-jet break}.

\begin{figure*}
	\includegraphics[width=\textwidth]{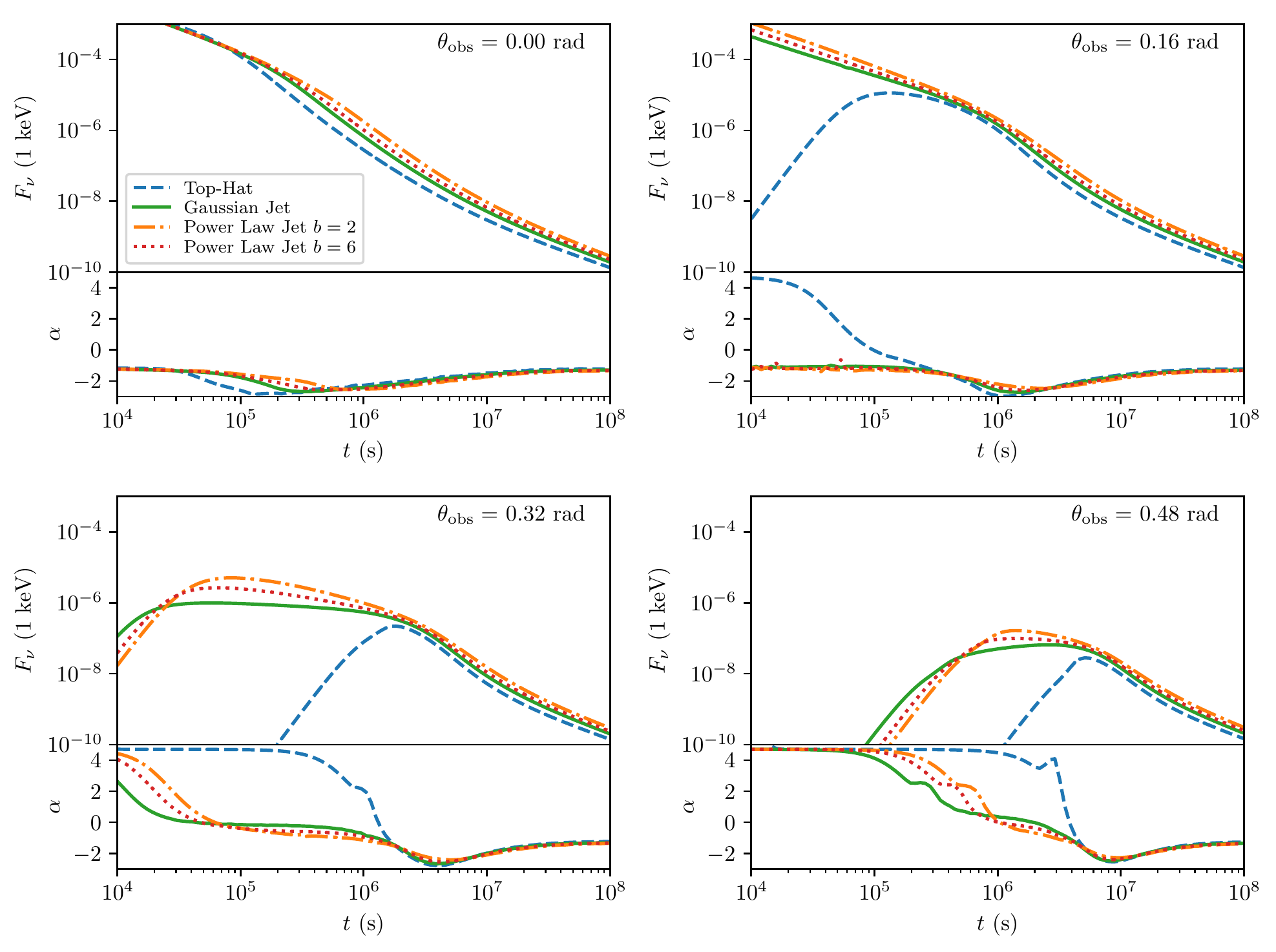}
	\caption{X-ray afterglow light curves from structured jets at different viewing angles $\thobs$, calculated with \afterglowpy{}.  Each panel shows the 1 keV flux density $F_\nu$ and the corresponding temporal slope $\alpha = d\log F_\nu / d\log \tobs$.  Included models are the top-hat jet (dashed blue), Gaussian jet (solid green), $b=2$ power law (dot-dashed orange), and $b=6$ power law at $\thobs=0$ (upper left), $\thobs=2\thC=0.16$ rad (upper right), $\thobs=4\thC=0.32$ rad (lower left), and $\thobs=6\thC=0.48$ rad (lower right). These particular light curves all use $\thC = 0.08$ rad, $\thW = 0.24$ rad, $E_0 = 10^{53}$ erg, $n_0=1$ cm$^{-3}$, $p=2.2$, $\epse = 10^{-1}$, $\epsB = 10^{-2}$, $d_L=10^{28}$ cm and $z=0.5454$. \label{fig:thVmodel}}
\end{figure*}

Figure \ref{fig:thVmodel} shows afterglow light curves from structured jet models at several viewing angles computed with \afterglowpy{}.  The upper left panel shows top-hat, Gaussian, $b=2$ power law, and $b=6$ power law jets viewed exactly on-axis at $\thobs=0$.  Each model shares $\thC=0.08$ rad, $\thW=0.24$ rad, $E_0=10^{53}$ erg, $n_0 = 1$cm$^{-3}$, $p=2.2$, $\epse=10^{-1}$, $\epsB = 10^{-2}$, $d_L=10^{28}$ cm and $z=0.5454$.  All of the models were computed with zero initial mass loading, so they do not show coasting phases.  They each show identical pre-jet break phases and similar post-jet break phases.  Each exhibits a slightly different jet break time, between $10^5$ s and $10^6$ s with these parameters, due to the different structure profiles $E(\theta)$.  This causes a discrepancy in the overall normalization, jets with more energetic wings but the same core with $\thC$ break later and remain brighter.  This discrepancy remains at all inclinations.

The upper right panel of Figure \ref{fig:thVmodel} shows the same models viewed at $\thobs=2\thC=0.16$ rad.  The top-hat jet exhibits an initial far off-axis phase which ends when the near side of the jet comes into view at $t=10^5$ s.  This initiates a prolonged transition period which completes  $\lesssim 10^6$ s as the far edge comes into view, leading into a normal post-jet break phase.  The structured models show similar pre-jet break phases, with slight normalization differences due to differing energies in the line of sight.  The light curves converge before the jet break at $t \lesssim 10^6$ s due to a brief structured phase in the Gaussian and $b=6$ models, which shallows their decay slightly.   All models show a jet break and transition to similar post-jet break phases after $10^6$ s.

The lower left panel of Figure \ref{fig:thVmodel} shows the same models viewed at $\thobs=4\thC=0.32$ rad.  Since $\thW=0.24$ rad all models show an initial far-off axis phase.  The structured models transition to a structured phase between $10^4$ s and $10^5$ s, much shallower than the standard pre-jet break behaviour.  The slope is shallowest in the the models with sharpest profiles: the Gaussian and $b=6$ power law.  All models show a jet break at $t\approx 2\times 10^6$ s and similar post-jet-break phases.

The lower right panel of Figure \ref{fig:thVmodel} shows the same models viewed at $\thobs=6\thC=0.64$ rad.  The far off axis phase lasts longer, and the structured phases have shallower slopes.  The Gaussian model even shows an increasing light curve.  Glitches in the $\alpha(\tobs)$ curves are due to the numerical treatment of jet spreading.  All models show a jet break at $t\approx 6\times 10^6$ s and similar post-jet-break phases.

It should be noted that the convergence of the light curves near the jet break is due in part to our standardized definition of $\thC$ in Equation \eqref{eq:thCdef}.  Alternative definitions, particularly omitting the factor of $b$ in the denominator of the power law profile expression, lead to larger differences in normalization with fewer shared features between light curves at the same $\thC$.  

\subsection{Structured Phase: Closure Relations}\label{subsec:structure}

Closure relations for the pre-jet break and post-jet break afterglow phases are well known in the literature (e.g. \citet{Granot:2002aa, Rhoads:1999aa}).  The closure relations for the far off axis phase in some spectral regimes first appeared in \citet{Salmonson:2003aa}.  Here we provide novel closure relations for the structured phase, as well as scalings for the jet break time with viewing angle.

\begin{figure}
	\includegraphics[width=\columnwidth]{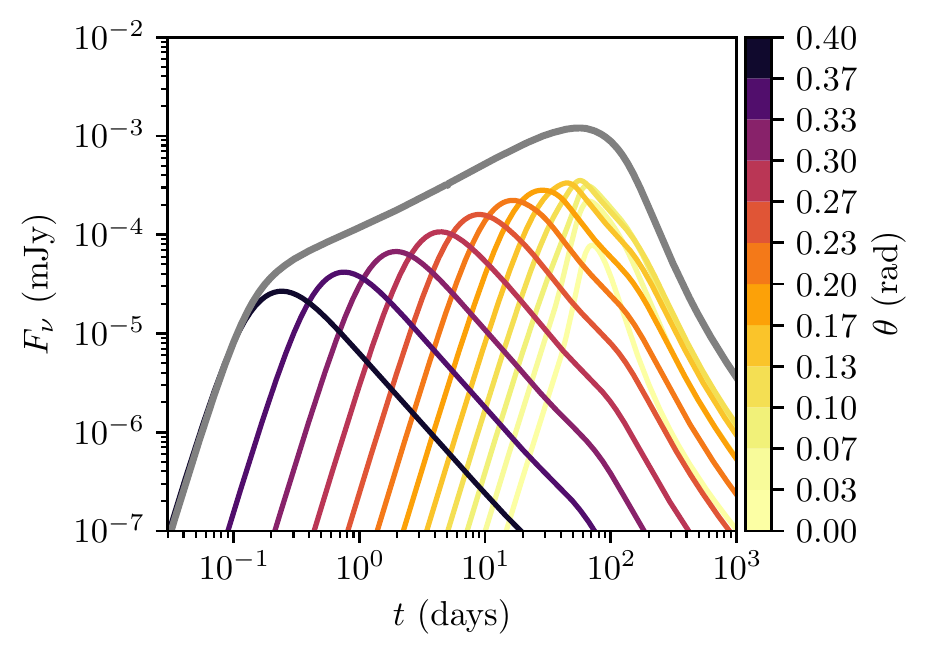}
	\caption{A Gaussian structured jet, decomposed into emission from different latitudes $\theta$.  The off-axis, structured, and post-jet break phases are all clearly visible.  The structured phase is the result of the brightest point of the blast-wave tracking from the wings $(\theta =\thW$) to the jet core ($\theta=\thC$). \label{fig:decomp}}
\end{figure}

Figure \ref{fig:decomp} shows the afterglow light curve of a Gaussian structured jet at viewed at $\thobs = 5 \thC = 0.5$ rad, clearly showing the far off axis, structured, and post jet break phases.  Also shown is a decomposition of the light curve into the contributions from each annular shell, from $\theta =\thW$ to $\theta = 0$.  The emission at any given time is dominated by material in a particular annular shell ($\theta = \theta_*$), which tracks towards the jet axis with time.  Material closer to the pole ($\theta < \theta_*$) has too high a Lorentz factor and is beaming suppressed, while material closer to the line of sight ($\theta > \theta_*$) has already peaked in emission and decreases with the standard pre-jet-break rate.  The material dominating the emission is that which just entered the beaming cone, with $\gamma_*^{-1} \sim \thobs-\theta_*$.

We can calculate the structured jet closure relations by determining how the time and magnitude of the peak of each annular section depends on the angle $\theta_*$.  The details are given in Appendices \ref{app:derive1} and \ref{app:derive2}; we give only the results here.  For brevity we define:
\begin{equation}
	\chi_* \equiv 2 \sin\left[\left(\thobs - \theta_*\right)/2\right] \approx \thobs-\theta_* \ . \label{eq:chiDef}
\end{equation} 
During the structured phase, the time at which the $\theta_*$ section peaks (dropping all pre-factors that do not depend on $\theta_*$) is:
\begin{equation}
	\tobs(\theta_*) \propto E(\theta_*)^{1/3} \chi_*^{8/3}\ . \label{eq:tobsTheta}
\end{equation}
While the blast wave is relativistic the rest-frame synchrotron emissivity has a power law dependence on the Lorentz factor and time in the burster frame: $\epsilon'_{\nu'} \propto \gamma_*^{s_\gamma} t^{s_t}$.  We can then write the total received flux from section $\theta_*$:
\begin{equation}
	F_\nu(\theta_*) \propto E(\theta_*)^{1 + s_t/3} \chi_*^{-s_\gamma + 2s_t/3 + \som + \beta} \nuobs^\beta\ .\label{eq:FnuTheta}
\end{equation}
Differentiating Equations \eqref{eq:tobsTheta} and \eqref{eq:FnuTheta} with respect to $\theta_*$ we find the slope of the light curve to be:
\begin{equation}
	\alpha \equiv \frac{d \log F_\nu}{d \log \tobs} = \frac{3 \beta - 3s_\gamma + 2s_t + 3\som + (3+s_t) g(\theta_*)}{8+g(\theta_*)}\ , \label{eq:structSlope}
\end{equation}
where
\begin{equation}
	g(\theta_*) \equiv -2\tan\left(\frac{\thobs-\theta_*}{2}\right) \left . \frac{d \log E}{d \theta}\right |_{\theta_*} \ . \label{eq:gdef}
\end{equation}
Equation \eqref{eq:structSlope} is the standard pre-jet break afterglow temporal slope, with dependence on $\som$ made explicit and the additional viewing angle dependent $g$ parameter.  The $g$ parameter is positive and to first order completely accounts for the presence of angular structure, a non-trivial $E(\theta)$, in the blast wave.  It is zero when the dominant material is in the line of sight ($\theta_* = \thobs$) or the jet is a top hat ($dE/d\theta = 0$).

\begin{figure}
	\includegraphics[width=\columnwidth]{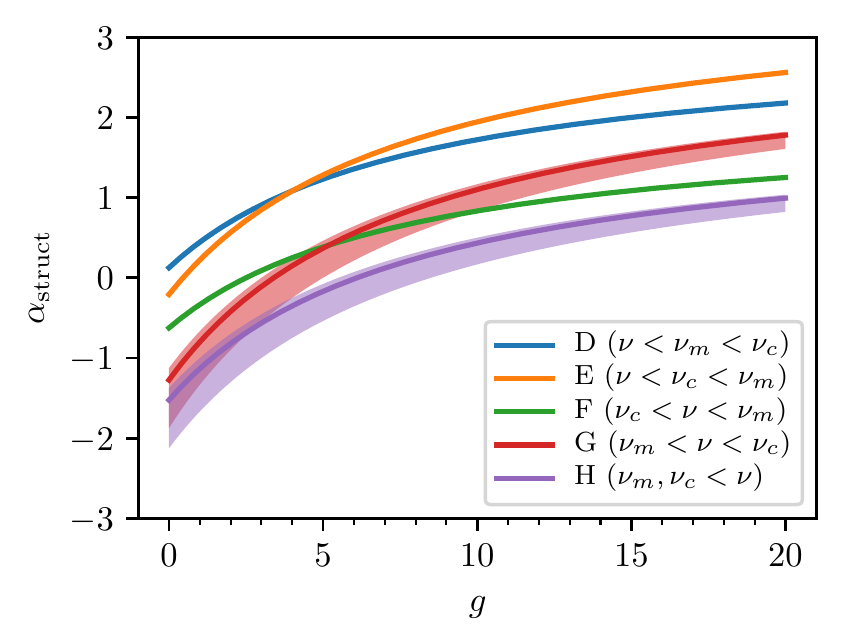}
	\caption{Temporal slopes $\alpha_{\mathrm{struct}}$ (defined as $F_\nu \propto \tobs^{\alpha}$) for the \emph{structured} phase as a function of the structure parameter $g$ for the synchrotron spectral regimes D (blue, narrow), E (orange, narrow),  F (green, narrow), G (red, wide), and H (purple, wide).  The width of the G and H bands shows the dependency on $p$, the upper limit for $p=2$, the lower for $p=3$, and the thick internal line for $p=2.2$. The temporal power law slope of a structured jet afterglow increases with $g$ in all spectral regimes. \label{fig:slopesG}}
\end{figure}

Figure \ref{fig:slopesG} shows the structured phase power law temporal slope $\alpha_{\mathrm{struct}}$ as a function of $g$ for spectral regimes $D$--$H$. In the $G$ and $H$ regimes the spectral slope $\beta$ is a function of $p$, this dependence carries over into the temporal slope as well. The temporal slope $\alpha_{\mathrm{struct}}$ is an increasing function of $g$ in all regimes.  At sufficiently large $g$ all regimes exhibit rising light curves, the specific $g$ at which the light curve begins to rise is regime dependent.

The $g$ parameter evolves with time as $\theta^*$ sweeps from the jet edge to the core.  This produces deviations in the light curve from a pure power law.  However, we find ultimately these deviations are not too large and the average slope is well approximated by $\geff  = g(\theff)$, where $\theff$ is a fiducial angular section which depends on viewing angle and potentially other structure parameters.  For the Gaussian model $\theff$ is simply half the viewing angle, for the power law model a suitable fitting function was found through comparison to a grid of \afterglowpy{} light curves:
\begin{align}
	\theff &= \frac{\thobs}{2}&& \text{Gaussian} \\
	\theff &= \thobs \big [1.8+2.1b^{-1.25}  && \text{power law} \nonumber  \\
	 	& \qquad + (0.49-0.86b^{-1.15}) \thobs/\thC \big ] ^{-1/2} &&\label{eq:theff}
\end{align}
In the expressions for $\geff$ it is reasonable to assume $2 \tan((\thobs-\theff)/2) \approx \thobs-\theff$ leading to the manageable expressions:
\begin{align}
	\geff &= \frac{\thobs^2}{4\thC^2} && \text{Gaussian}\ , \\
	\geff &= \frac{2 b (\thobs-\theff)\theff}{b \thC^2+\theff^2} && \text{power law}\ . \label{eq:geff}
\end{align}
These values may be used in Table \ref{tab:slopes} for an approximate power law model of a structured jet with an appropriate viewing angle dependent temporal evolution.  

\begin{figure*}
	\includegraphics[width=\textwidth]{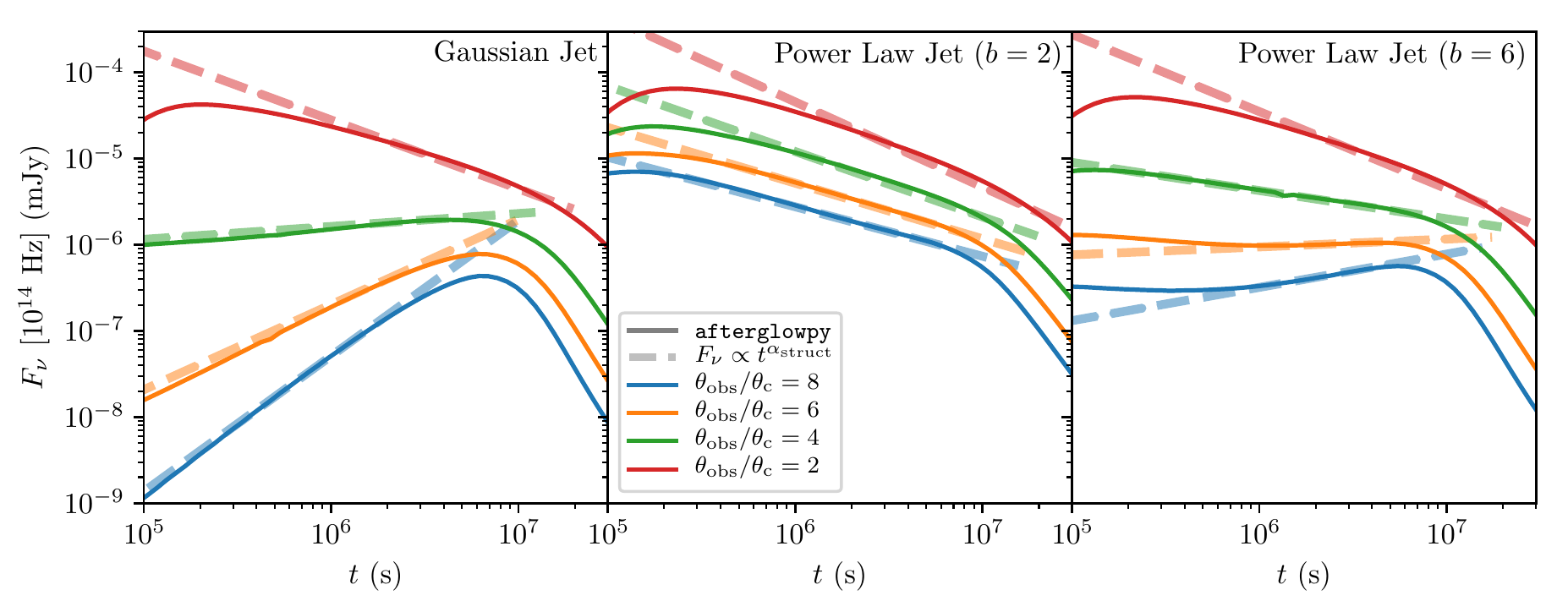}
	\caption{Optical afterglow light curves in the \emph{structured} phase with corresponding closure relation slopes $\alpha_{\mathrm{struct}}$.  Each panel shows light curves with $\thobs/\thC$ = 2, 4, 6, and 8 (solid lines in red, green, orange, and blue respectively) calculated with \afterglowpy{} and corresponding power law segments $F_\nu \propto \tobs^{\alpha_{\mathrm{struct}}}$ calculated from the closure relations in Table \ref{tab:slopes} and the $\geff$ values from Equation \eqref{eq:geff}.  The left panel shows a Gaussian jet, the middle a $b=2$ power law jet, and the right a $b=6$ power law jet.  The closure relations $\alpha_{\mathrm{struct}}$ capture the approximate power law slope of the numerical light curves, with the agreement improving for larger values of $\thobs/\thC$. The light curves use a fixed $\thobs$ = 0.3 rad, $\thW = 0.25$ rad, and vary $\thC$.  Other parameters are fixed: $E_0 = 10^{53}$ erg, $n_0=10^{-3}$ cm$^{-3}$, $p=2.2$, $\epse = 10^{-1}$, $\epsB = 10^{-3}$, $d_L=10^{28}$ cm and $z=0.5454$. \label{fig:closure}}
\end{figure*}

Figure \ref{fig:closure} shows \afterglowpy{} light curves for Gaussian, $b=2$ power law, and $b=6$ power law jets at various values of $\thobs / \thC$ and the corresponding $\alpha_{\mathrm{struct}}$ slopes calculated using $\geff$.  The analytic closure relations capture the light curve behaviour in the structured phase for all 3 models, with somewhat better agreement at larger values of $\thobs/\thC$.  The Gaussian jet, with its strong energy gradient, is most sensitive to changes in $\thobs/\thC$.  The $b=6$ power law, with its more energetic wings, displays less sensitivity to changes in $\thobs/\thC$.  The $b=2$ power law displays almost no variation at all in the temporal decay rate.  The overall behaviour of the Gaussian and $b=2$ power law jets are in agreement with \citet{Rossi:2004aa} (R04, see for comparison their Figure 12 and 16).  Their $\thobs/\thC = 4$ Gaussian jet showed a mildly decaying slope instead of the rising light curve seen in Figure \ref{fig:closure}, due to the choice of different values of $p$.  If $\thobs/\thC=4$, then a Gaussian jet will have also $\geff = 4$ by Equation \eqref{eq:geff}, and Figure \ref{fig:slopesG} shows $\alpha_{\mathrm{struct}}$ will be negative for $p=2.5$ in the G regime as in R04.

In the structured phase, the temporal slope depends directly on $\geff$ in a way independent of the particular jet model.  If a particular structure is assumed, then $\thobs/\thC$ may easily be inferred from $\geff$, and hence from $\alpha_{\mathrm{struct}}$.

\subsection{Structured Phase: Break Times}\label{subsec:structureBreak}

The break times, the times when an afterglow light curve transitions from one phase to another, scale with the blast wave parameters in a well-defined way. Firstly, a convenient characteristic time scale is the non-relativistic time $\tNR$, the time at which the core of the jet would become non-relativistic in the absence of spreading:
\begin{align}
	\tNR &= (1+z) \left( \frac{9}{16\pi} \frac{E_0}{\rho_0 c^5} \right)^{1/3}  \label{eq:tNRdef} \\
	&= 882 \ (1+z) E_{53}^{1/3}n_0^{-1/3} \mathrm{ days}\ . \nonumber
\end{align}
In Equation \eqref{eq:tNRdef} $E_{53} = E_0 / (10^{53}$ erg) and $n_0$ is the circumburst density measured in cm$^{-3}$.  While the jet is relativistic any time scale will depend on $E_0$ and $n_0$ only through $\tNR$.

For viewers outside the truncation angle, $\thobs > \thW$, the transition from the far-off-axis phase to the structured phase occurs at $\tW$ (see rightmost panel of Figure \ref{fig:phaseSketch}).  The transition marks when the observers enters the beaming cone of the near edge of the jet: $\gamma^{-1}(\thW) \sim \thobs-\thW$.  This will occur at observer time:
\begin{align}
	\tW &= \tNR \left(E(\thW) / E_0\right)^{1/3} \left( \thobs-\thW \right)^{8/3}  \label{eq:tw} \\
	&= 12.1 \ (1+z) E_{53}^{1/3} n_0^{-1/3} \nonumber \\
	& \qquad \times \left(\frac{E(\thW) }{ E_0}\right)^{1/3} \left( \frac{\thobs-\thW}{0.2\ \mathrm{ rad}} \right)^{8/3}  \mathrm{ days}\ .
\end{align}
The factor of $(E(\thW) / E_0)^{1/3}$ in $\tW$ compensates for the lower Lorentz factor (and hence larger beaming cone) in the wings of the jet relative to the core.  It depends only on geometric parameters (e.g. $\thC$, $\thW$, and $b$) and may be much smaller than 1.

The \emph{jet break} $\tb$ is the time at which the effective angular size of the jet ceases to increase: $\som = 0$.  For an on-axis observer of a top hat jet, in the absence of spreading, this is when the jet edges come into view.  More generally, this is the moment when the blast wave has sufficiently decelerated that the viewer is within the beaming cones of its brightest regions.  For \emph{aligned} viewers, $\thobs < \thC$, the jet break marks the transition from pre- to post-jet break phases.  For \emph{misaligned} viewers, $\thobs > \thC$, the jet break marks the transition from structured to post-jet break phases.

To determine the normalization and scaling for the observed jet break time $\tb$ we computed a grid of afterglow light curves for Gaussian, $b=2$ power law, and $b=6$ power law jets, varying $\thC$ and $\thobs$.  We used 11 values each for $\thC$ between 0.04 rad and 0.40 rad and $\thobs$ between 0 rad and 1.0 rad.  The truncation angle $\thW$ was set to $10 \thC$, and all other parameters were fixed.  Each light curve was fit with a smoothly broken power law $F_\nu \propto (\tobs/\tb)^{\alpha_1} [(1+(\tobs/\tb)^s)/2]^{(\alpha_2-\alpha_1)/s}$ in the region surrounding the jet break to identify the precise break time $\tb$.

\begin{figure*}
	\plottwo{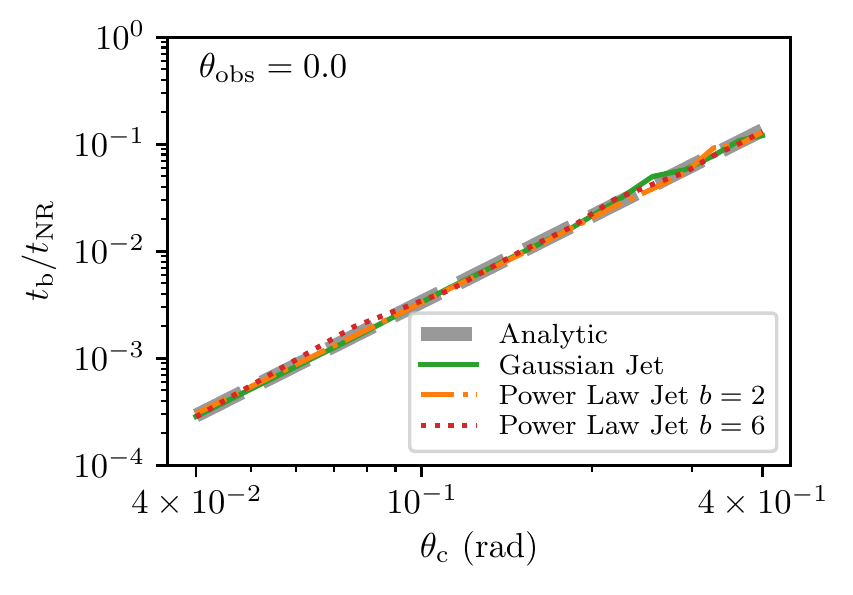}{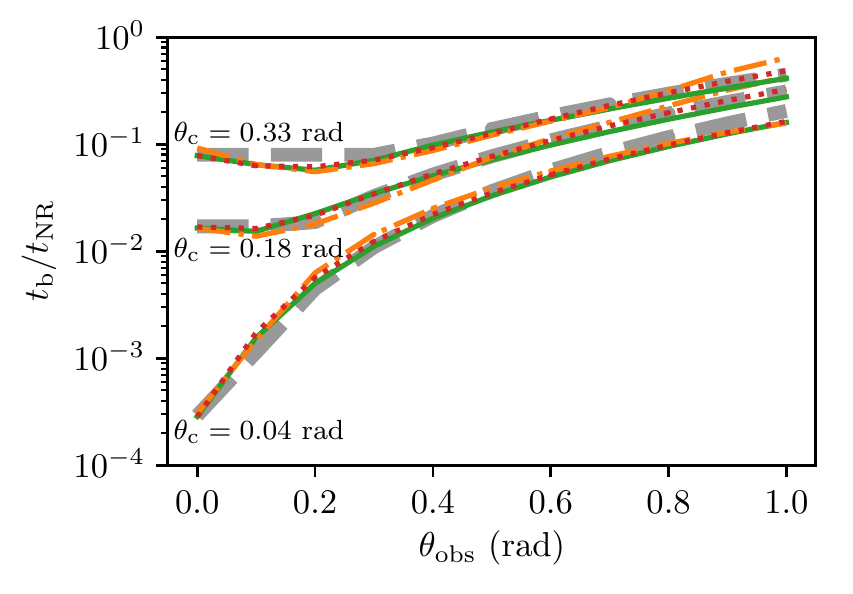}
	\caption{Jet break time $t_b$ for Gaussian (green solid line), $b=2$ power law (orange dash-dotted line), and $b=6$ power law (red dotted line) jets.  The left panel shows $t_b$ as a function of $\thC$ for on-axis (\thobs=0) observers.  The right panel shows $t_b$ as a function of $\thobs$: the lower, middle, and upper curves correspond to $\thC = $ 0.04 rad, 0.184 rad, and 0.328 rad respectively. The dashed grey line is the analytic approximation Equation \eqref{eq:tb}. Each break time was determined by fitting a smoothly broken power law to an \afterglowpy{} light curve. \label{fig:tb}}
\end{figure*}

Figure \ref{fig:tb} shows the extracted jet break times from \afterglowpy{} light curves for several jet structure models.  The left panel shows $\tb$ as a function of $\thC$ for on-axis ($\thobs=0$) viewers, while the right panel shows $\tb$ as a function of $\thobs$ for 3 particular values of $\thC$.  The scale for $\tb$ has been normalized by $\tNR$, isolating the variation due to structure geometry and viewing angle.  The different structure models show remarkably tight agreement, particularly as $\thobs \to 0$.  This is primarily due to the shared definition of $\thC$.  As $\thobs \to 1$ rad the power law jets show a slight increase relative to the Gaussian jet, particularly for the larger $\thC$.  

It has been long established that on-axis viewers observe the jet break in their frame at a time $\tb \propto \thC^{8/3}$ (e.g. \citet{Rhoads:1999aa}), while significantly off-axis viewers observe the jet break at $\tb \propto \thobs^{8/3}$ \citep{Rossi:2002aa, Panaitescu:2003aa}.  More generally, for initially top hat jets it has been shown the jet break corresponds to the far jet edge coming into view, that is $\tb \propto (\thC + \thobs)^{8/3}$ \citep{van-Eerten:2010aa}.  As $\thC$ denotes the angular size of the bright jet core, once this time is reached the jet is essentially entirely in view.  Wings of the jet on the far side from the viewer may still enter the beaming cone, but remain a subdominant contribution to the bright core within $\thC$.  

For generic structured jets we find $\tb(\thC, \thobs)$ to show two modes of behaviour, split by the whether the viewer is aligned ($\thobs < \thC$) or misaligned ($\thobs > \thC$).  Aligned viewing shows $\tb$ to follow $\thC^{8/3}$ quite strongly with very little dependence on $\thobs$.  Misaligned viewing better tracks with a linear combination $(\thobs + c_1 \thC)^{8/3}$.  To determine the precise normalization we fit each regime with a simple $\chi^2$ minimization.

We refer to the jet break time in the aligned case as $\tbin$.  It is fit by:
\begin{align}
	\tbin &= 1.56\ \tNR \thC^{8/3} \label{eq:tbin}\\
		&= 2.95\ E_{53}^{1/3} n_0^{-1/3} \left(\frac{\thC}{0.1\ \mathrm{rad}}\right)^{8/3}\ \mathrm{days}\ . \nonumber 
\end{align}
The jet break time in the misaligned case is $\tbout$, and fit by:
\begin{align}
	\tbout &= 0.180\ \tNR (\thobs+1.24\ \!\thC)^{8/3} \label{eq:tbout} \\
		&= 24.9\ E_{53}^{1/3} n_0^{-1/3} \left(\frac{\thobs+1.24\ \!\thC}{0.5\ \mathrm{rad}}\right)^{8/3}\ \mathrm{days}\ . \nonumber 			
\end{align}

A good fit for $\tb$ to the full grid of models is obtained piecewise by switching from $\tbin$ to $\tbout$ at the transition point $\thobs = 1.01 \thC$ where $\tbin = \tbout$.  That is:
\begin{equation}
	\tb = \begin{cases}
			\tbin & \thobs < 1.01 \thC \\
			\tbout & \thobs > 1.01 \thC
		\end{cases} \ . \label{eq:tb}
\end{equation}

Figure \ref{fig:tb} also shows Equation \eqref{eq:tb} compared to the measured jet break times.  The on-axis relation $\tbin$ fits very well over the span of $\thC$ considered, with a mean fractional error of 5.3\%.  Over the full $\thC$ and $\thobs$ domain the mean fractional error is 12\%, with the largest contributions coming from large viewing angles $\thobs > 0.8$ rad.

\subsection{Refreshed Shocks --- Fast Tails and Quasi-Spherical Cocoons}\label{subsec:refreshedShocks}

An alternative mechanism to produce slow decays or rises in afterglow light curves is the \emph{refreshed shock}, where velocity stratification of the ejecta causes a prolonged period of energy injection in the afterglow \citep{Rees:1998aa, Panaitescu:1998aa, Sari:2000aa}.  This material may be a tail of fast outflowing ejecta or ejecta material accelerated through interaction with a possibly choked jet (e.g. a cocoon) \citep{Nakar:2011aa, Hotokezaka:2015aa}.  Both scenarios are particularly relevant to afterglows arising from binary neutron star mergers.  In either case, slow material initially coasting behind the shock is gradually incorporated into the blast wave as it decelerates, ``refreshing'' the shock.  This mechanism is a general scenario for energy injection \citep{Zhang:2006aa}, and was proposed as the mechanism behind the \gwbns{} afterglow in the choked-jet picture (e.g. \citep{Mooley:2018aa, Nakar:2018aa}).

In the simplest case the visible part of the blast wave is assumed to be quasi-spherical.  The velocity distribution of material behind the blast wave is specified by $E(u)$, the energy of all material in the ejecta with four-velocity greater than $u$.  This is typically taken to be a power law in $u$ within the finite domain $[\umin, \umax]$:
\begin{equation}
	E(u) = E_0 \left(\frac{u}{\umax}\right)^{-k}\ , \label{eq:Eu}
\end{equation}
where $u \in (\umin, \umax)$ is the dimensionless four-velocity, $k>0$ is the power law index, and $E_0$ the kinetic energy of the fastest material with $u=\umax$ initially.  The mass ejected with velocity $\umax$ is $\Mej = E_0 / ((\gmax -1)c^2)$.

A blast wave will have an initial coasting period before sweeping up enough mass in the ambient medium to begin deceleration and be subject to refreshed shocks.  The transition between coasting and decelerating occurs in the bursters frame at $\tdec$. The shock refreshment ends at $\tumin$ when the blast wave decelerates past $\umin$, when there is no longer any material to refresh it.   At this point the afterglow transitions to a standard pre-jet break phase. In the single shell model with in the limit $\umax \gg 1$, these transition times (in the observer's frame) are:
\begin{align}
	\tdec &= \frac{1}{4}(1+z)   \umax^{-8/3}\ \! \tNR  \ , \label{eq:tdec} \\ %\left(\frac{9}{4\pi} \frac{\Mej}{n_0 \Mp c^3} \frac{\gmax^2}{\bmax^3(\gmax+1)(4\umax^2+3)} \right)^{1/3}  \ ,\label{eq:tdec} \\
	\tumin &= \frac{1}{4}(1+z) \frac{2+k}{8+k} \left(\frac{\umax}{\umin}\right)^{(8+k)/3} \umax^{-8/3}\ \! \tNR \ .\label{eq:tumin}
\end{align}
 The temporal and spectral slopes for the coasting and refreshing phases are given in Table \ref{tab:refreshSlopes}.  The refreshing phase slope $\alpha_{\mathrm{refresh}}$ depends on the energy index $k$.  A quasi-spherical blast wave experiences very little or no jet spreading and should smoothly transition to the Newtonian Sedov phase after $\tumin$ passes and the energy injection ends.

\begin{deluxetable}{CCCCC}
	\tablecaption{Refreshed shock temporal and spectral slopes: $F_\nu \propto \tobs^{\alpha} \nu^{\beta}$.  See Equation \eqref{eq:Eu} for the definition of $k$.  \label{tab:refreshSlopes}}
	\tablehead{\colhead{Regime}& \colhead{Label} & \colhead{$\alpha_{\mathrm{coast}}$}& \colhead{$\alpha_{\mathrm{refresh}}$} & \colhead{$\beta$}}
	\startdata
	\nu<\nu_m<\nu_c     & D & 3 	& \frac{4 + 3k}{8+k} 		& 1/3 \\
	\nu < \nu_c < \nu_m & E & 11/3 	& \frac{4/3 + 11k/3}{8+k} 	& 1/3 \\
	\nu_c < \nu < \nu_m & F &  2 	& \frac{-2 + 2k}{8+k} 		& -1/2 \\
	\nu_m<\nu<\nu_c     & G & 3 	& \frac{-6p + 6 + 3k}{8+k} 	& -(p-1)/2 \\
	\nu_c < \nu_m < \nu & H & 2 	& \frac{-6p + 4 + 2k}{8+k}	& -p/2 \\ 
	\enddata
\end{deluxetable}

\subsection{Relation to Energy Injection Models}

The light curves of both the misaligned angularly structured jet and refreshed shock models can be understood in terms of general energy injection mechanisms, parameterized as an additional power $L$ delivered to the blast wave.  This power is typically taken to decay in time as a power law, $L \propto t^{-q}$, with $q \in [0,1]$.

The mapping between energy injection and refreshed shocks was done by \cite{Zhang:2006aa}.  That work specifies the index $s$ of the mass distribution $M_{>u} \propto u^{-s}$, which is related simply to the index $k$ by $k = s-1$. Under this mapping the closure relations in Table \ref{tab:refreshSlopes} are equivalent to those in \cite{Zhang:2006aa} and the model can be seen as equivalent to an energy injection model with $q = (8-2k) / (8+k)$.

Misaligned structured jets do not map as neatly onto energy injection models as the scaling of the visible patch, $\som = 1.0$, is different than in standard afterglows, $\som = 2$. Once this change is accounted for, the $g$ parameter plays the same role as $k$, as can be seen by a simple comparison between $\alpha_{\text{struct}}$ in Table \ref{tab:slopes} and $\alpha_{\text{refresh}}$ in Table \ref{tab:refreshSlopes}.  That is, a structured jet light curve is similar to a standard energy injection model with:
\begin{align}
	q &= (8-2g) / (8+g)\ ,  \label{eq:struct2ei} \\
	\som &= 1\ .
\end{align}

It should be noted that although one can map structured jet light curves onto an energy injection model, there is no energy being added to the blast wave. Rather, the ``injection'' of energy is the additional flux from higher energy portions of the jet decelerating to the point the observer is in their beaming cone.

%%%%%%%%%%%%%%%%%%%%%%%
%
%  GW170817A
%
%%%%%%%%%%%%%%%%%%%%%%%%%

\section{Application: \gwbns{}}\label{sec:gw170817}

We use the electromagnetic afterglow of \gwbns{} as an example application of both the analytic and numerical structured jet afterglow models, using the dataset from \citet{Troja:2019ab} with additional \hubble{} observations reported in \citet{Lamb:2019aa}. No new observations are presented in this paper.

\begin{figure*}
	\includegraphics[width=\textwidth]{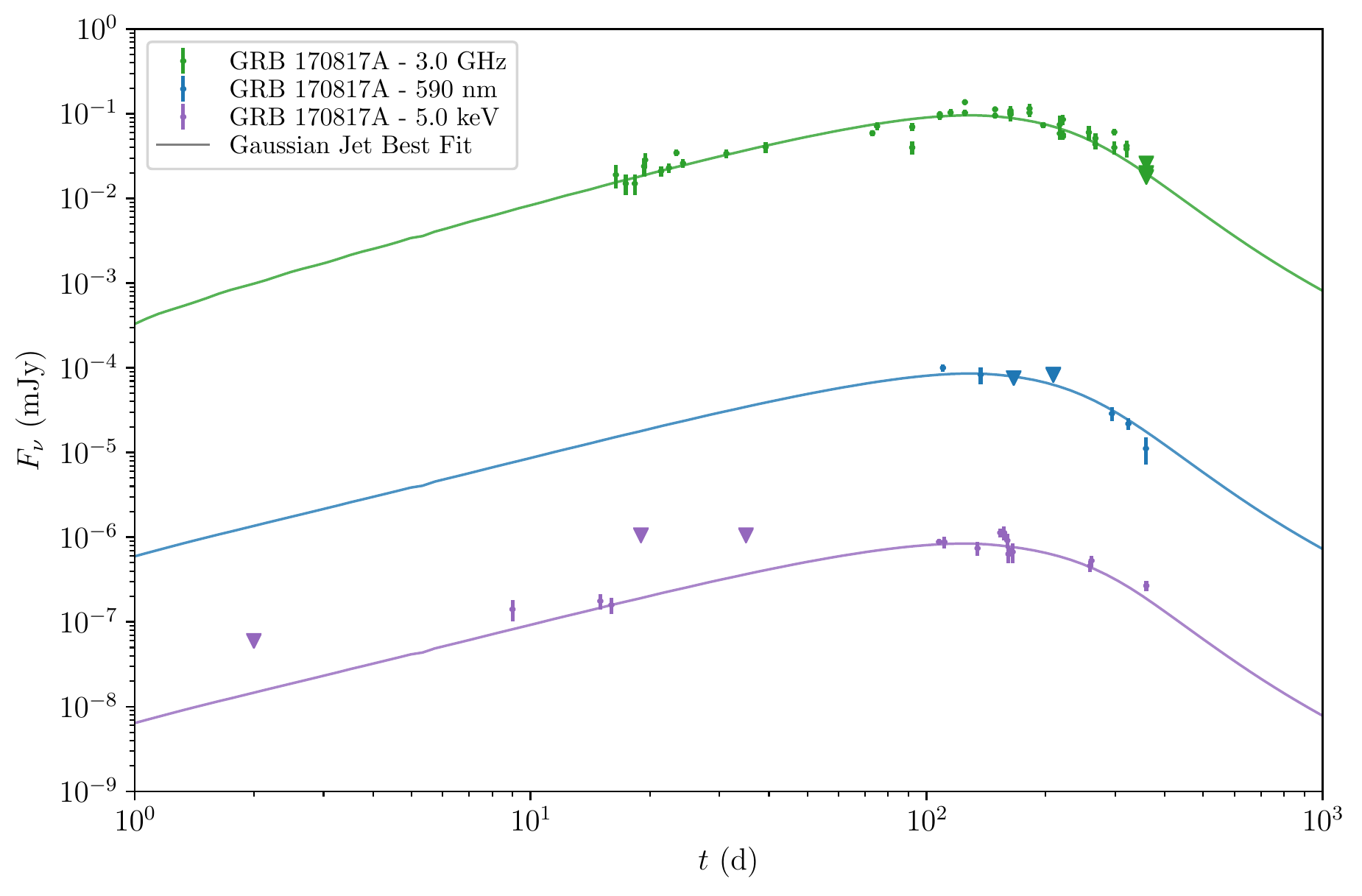}
	\caption{The afterglow light curve of \grbbns{} in radio (3 GHz, green, top), optical (590 nm, blue, middle), and X-ray (5 keV, purple, bottom).  For illustrative purposes, radio and optical data have been shifted to their fiducial frequencies assuming $F_\nu \propto \nu^{-0.585}$, appropriate for the G synchrotron regime with $p=2.17$. Downward pointing triangles are 3-sigma upper limits.  Solid lines show the best fit (maximum posterior probability) Gaussian jet light curve.  Some upper limits with low constraining power were not included in the figure.  The dataset is that of \citet{Troja:2019ab} with additional \hubble{} data points from \citet{Lamb:2019aa}.  \label{fig:lc170817A}}
\end{figure*}

Figure \ref{fig:cornerGaussian} shows the observations in radio, optical, and X-ray bands of the \gwbns{} afterglow.  No afterglow emission was detected until an X-ray detection by the Chandra observatory nine days after the GW trigger \cite{Troja:2017aa}.  Further radio, optical, and X-ray observations showed a steady rising light curve with constant power law slope until a peak near day 164 \citep{Haggard:2017aa, Hallinan:2017aa, DAvanzo:2018aa,  Lyman:2018aa, Margutti:2018aa, Mooley:2018aa, Troja:2018aa, Troja:2019ab}.  At this point the light curve turned over and began a steep decay that continues to the present \citep{Alexander:2018aa, Fong:2019aa, Lamb:2019aa, Troja:2019ab}.

Interpreting the afterglow as arising from a structured jet allows one to draw the following quick conclusions:
\begin{enumerate}
	\item The spectral slope $\beta = 0.585 \pm 0.005$ throughout the entire evolution, identifying the relevant spectral regime as G: $\nu_m < \nuobs < \nu_c$, and $p = 2.17 \pm 0.01$ \citep{Troja:2019ab}.
	\item The initial rising ($\alpha > 0$) light curve with declining spectrum ($\beta < 0$) immediately identify this as a \emph{misaligned} evolution ($\thobs > \thC$).  Aligned jets may only have a rising light curve when viewed at frequencies below the peak frequency, that is when $\beta < 0$.
	\item The rising light curve maintains $\alpha  = 0.90 \pm 0.06$ \citep{Troja:2019ab}.  Using the closure relations from Table \ref{tab:slopes} we can identify this as the \emph{structured} phase and determine $\geff = 8.2 \pm 0.5$.  If the light curve underwent an \emph{far-off-axis} phase, it concluded before the afterglow was detected: $\tW < 9$d.  The peak can be identified as the jet-break: $\tb = 164 \pm 12$ d.
	\item In the context of a Gaussian jet, the measurement of $\geff$ leads immediately to $\thobs / \thC = 5.7 \pm 0.2$ using Equation \eqref{eq:geff}.  A $b=2$ power law jet cannot achieve this $\geff$ for any reasonable value of $\thobs / \thC$.  A $b=6$ power law jet requires $\thobs/\thC = 14 \pm 1$, while a $b=9$ power law requires $\thobs/\thC = 9.6\pm 0.5$.
\end{enumerate}
To go further we first need to assume a value for $E_0 / n_0$. For demonstration we use the \citet{Fong:2015aa} fiducial values of $E_0 \sim 2\times 10^{51}$erg and $n_0 \sim 10^{-2}$cm$^{-3}$, assuming 1 dex uncertainty on both parameters.
\begin{enumerate}
	\setcounter{enumi}{4}
	\item We can use the jet break $\tb$ measurement to determine $\thobs + \thC$ using Equation \eqref{eq:tbout}. Using our assumed values for  $E_0$ and $n_0$ leads to $\thobs + 1.24 \thC \approx 0.93 \pm 0.38$ rad ($53^\circ \pm 22^\circ$). This does not depend on the jet model.
	\item For a Gaussian jet the combined measurements of $\thobs +1.24\thC$ and $\thobs/\thC$ then imply $\thC \approx 0.13 \pm 0.06$ rad ($7.6^\circ  \pm 3.1^\circ$) and $\thobs \approx 0.78 \pm 0.31$ rad ($44^\circ \pm 18^\circ$).
	\item For a $b=6$ power law jet we get $\thC \approx 0.062 \pm 0.026$ rad ($3.5^\circ  \pm 1.5^\circ$) and $\thobs \approx 0.85 \pm 0.35$ rad ($49^\circ \pm 20^\circ$).
\end{enumerate}

The relative weak dependence of $\tb$ on $E_0/n_0$ allows one to make estimates of $\thobs$ and $\thC$ even if $E_0$ and $n_0$ are only known to an order of magnitude. 

 Analysis of the \gwbns{} gravitational wave signal produced estimates of the inclination angle $\iota$ between the binary neutron stars' orbital plane and our line of sight, but were dependent on the assumed value of the Hubble constant $H_0$ \citep{Abbott:2017aa}.  The values obtained were:
 $\iota = \left({28}^{+27}_{-20}\right)^\circ$ (90\% uncertainties, marginalized over $H_0$), $\iota = \left(26^{+12}_{-15}\right)^\circ$ ($H_0$ from SHoES: \citet{Riess:2016aa}), and $\iota = \left(19^{+12}_{-12}\right)^\circ$ ($H_0$ from \citet{Planck-Collaboration:2016aa}), we have reversed the reported orientation of $\iota$ to ease comparison with $\thobs$.
 
If the BNS merger produces a jet orthogonal to the orbital plane then $\iota = \thobs$ and we can directly compare the gravitational and electromagnetic estimates.  With our fiducial values for $E_0$ and $n_0$ the estimated value of $\thobs$ from the afterglow is larger than the gravitational wave $\iota$ measurements, but given the large uncertainties on both values there is no tension in the discrepancy.

There is much more information in the afterglow light curve than just the rising slope and jet break.  For a more detailed analysis we performed Bayesian parameter estimation utilizing \afterglowpy{} and the \emcee{} Markov chain Monte Carlo sampler \citep{Foreman-Mackey:2013aa}.  We used the light curve data reported in \citet{Troja:2019ab} with additional \hubble{} observations reported in \citet{Lamb:2019aa}. Reported measurement uncertainties were assumed to be independent and Gaussian, upper limits were treated as observations of zero flux with a 1-sigma Gaussian uncertainty. We assume $\iota = \thobs$ and use the posterior probability distribution $p_{\mathrm{LIGO}}(\cos \iota)$ reported in \citet{Abbott:2017aa} (assuming Planck $H_0$, their Figure 3) as a prior for $\thobs$ as in \citet{Troja:2018aa}.

We fit both the Gaussian and power law structured jet models, fixing $\xi_N = 1$ and $\dL = 1.23 \times 10^{26}$ cm.  The fit parameters, their priors, and their marginalized posteriors are given in Table \ref{tab:MCMC}. Both models were run with 300 walkers for 64 000 iterations, discarding the first 16 000 iterations as a burn-in.  The $E_0$, $n_0$, $\epse$, and $\epsB$ parameters were given $\log$-uniform priors, while $\thC$, $\thW$, $b$, and $p$ were given uniform priors.  The specific bounds and priors for each parameter are reported in Table \ref{tab:MCMC}.

\begin{deluxetable*}{CCCCCC}
	\tablecaption{Parameter estimation priors and marginalized posteriors for the \gwbns{} afterglow using the \afterglowpy{} Gaussian and power law jet models, including viewing angle constraints from LIGO assuming the \planck{} value of $H_0$.  Given posterior values for each model are the median, 16\%, and 84\% quantiles.  Parameters in the lower section are derived from the posterior distributions of the fit parameters in the upper sections. \label{tab:MCMC}}
	\tablehead{\colhead{Parameter}& \colhead{Unit} & \colhead{Prior Form} & \colhead{Bounds} & \colhead{Gaussian Jet Posterior} & \colhead{Power Law Jet Posterior}}
	\startdata
		\thobs	& \text{rad	} & \sin \thobs \times p_\mathrm{LIGO}(\cos \thobs)	& [0, 0.8] 	& 0.40^{+0.11}_{-0.11} 	& 0.44^{+0.12}_{-0.12} \\[0.05cm]
		\log_{10} E_0	& \text{erg	}	& \text{uniform}	& [45, 57] 						& 52.96^{+0.97}_{-0.72} 	& 52.93^{+1.1}_{-0.75}\\[0.05cm]
		\thC			& \text{rad	}	& \text{uniform}	& [0.01, $\pi/2$]					& 0.066^{+0.018}_{-0.018} & 0.046^{+0.013}_{-0.013}\\[0.05cm]
		\thW			& \text{rad}	& \text{uniform}	& [0.01, 12$\thC$]				& 0.47^{+0.26}_{-0.19} 	& 0.238^{+0.071}_{-0.69}\\[0.05cm]
		b			& - 			& \text{uniform}	& [0, 10] 						& - 					& 9.03^{+0.70}_{-1.1}\\[0.05cm]
		\log_{10}n_0	& \text{cm}^{-3} & \text{uniform}	& [-10, 10] 					& -2.70^{+0.95}_{-1.0} 	& -2.6^{+1.1}_{-1.1}\\[0.05cm]
		p			& - 			& \text{uniform}	& [2, 5] 						& 2.168^{+0.063}_{-0.0075} & 2.1653^{+0.0085}_{-0.010}\\
		\log_{10}\epse& - 			& \text{uniform}	& [-5, 0] 						& -1.42^{+0.70}_{-1.1} 	&-1.24^{+0.73}_{-1.2} \\[0.05cm]
		\log_{10}\epsB	& - 			& \text{uniform}	& [-5, 0] 						& -3.96^{+1.1}_{-0.74} 	& -3.76^{+1.1}_{-0.87}\\[0.05cm]
		\hline
		\log_{10}\Etot	& \text{erg}	&	-		&	-						& 50.57^{+0.92}_{-0.66} 	&50.46^{+1.1}_{-0.73} \\[0.05cm]
		\thobs/\thC	& -			&	-		&	-						& 6.12^{+0.18}_{-0.18} 	& 9.38^{+0.73}_{-0.56}\\[0.05cm]
		\log_{10} E_0/n_0 &  \text{erg cm}^3	&	- 	&	-						& 55.69^{+1.1}_{-0.85} 	& 55.62^{+1.2}_{-0.83}\\[0.05cm]
	\enddata
\end{deluxetable*}

Table \ref{tab:MCMC} gives the posterior median and 68\% quantiles found for each parameter, as well as constraints on the total jet energy \Etot{} and the ratio $\thobs/\thC$. Figures \ref{fig:cornerGaussian} and \ref{fig:cornerPowerlaw} show the one- and two-dimensional marginalized views of the posterior distribution for the Gaussian and power law jet model fits, respectively.  The parameters $E_0$, $n_0$, $\epse$, and $\epsB$ are only constrained to within an order of magnitude, due partially to the observed radio, optical, and X-ray data all lying on the same synchrotron power-law segment.  These parameters are shared between jet models, and are consistent between the two fits.  The electron energy index $p$ is extremely well constrained by both models, of course, due precisely to the large range of data laying on the same synchrotron segment.  The total jet energy in both models is constrained to be on the order $3 \times 10^{50}$ erg, with uncertainties of an order of magnitude.

Of most interest are the geometric parameters: the viewing angle $\thobs$ and the jet structure parameters $\thC$, $\thW$, and $b$. Both models constrain $\thobs$ and $\thC$ reasonably well, $\thobs = 0.40 \pm 0.11$ rad for the Gaussian jet and $\thobs = 0.44 \pm 0.12$ rad for the power law, but constrain the combination $\thobs/\thC$ far better: $\thobs/\thC  = 6.12 \pm 0.18$ for the Gaussian and $\thobs/\thC  = 9.38^{+0.73}_{-0.56}$ for the power law. This is very evident by the corresponding map of the posterior in Figures \ref{fig:cornerGaussian} and \ref{fig:cornerPowerlaw}, which clearly displays a preferred linear relationship between $\thobs$ and $\thC$. This is the manifestation of the structured jet closure relations with $g \approx \geff(\thobs/\thC)$.  The looser constraint on $\thobs/\thC$ in the power law jet comes from the additional freedom of the $b$ parameter, which has an effect on $\geff$ in that model (see Equation \eqref{eq:geff}). The truncation angle $\thW$ is essentially unconstrained in the Gaussian fit as a far off-axis phase was not observed.  In the power law fit, however, $\thW$ is constrained to be quite narrow so as to avoid the early bright wings.

The fits agree with the back-of-the-envelope reasoning from the analytic closure relations and jet break time.  The closest agreement is in the $\thobs/\thC$ ratio, the Gaussian agrees within the uncertainties and the power law fit agrees well with the $b=9$ estimate. The absolute values of $\thobs$ and $\thC$ agree within uncertainties but are systematically higher in the back-of-the-envelope estimates compared to the fit results.  This can be accounted for by our adopted fiducial value for $E_0/n_0$ which in the estimate was taken to be $10^{53.3}$ erg cm$^3$.  The MCMC fits, which constrain this value for this particular burst, both find $E_0/n_0 \approx 10^{55.6 \pm 1.0}$ erg cm$^{3}$.  Adopting this value in the back-of-the-envelope estimates leads to $\thobs = 0.41\pm0.12$ rad, $\thC = 0.07 \pm 0.02$ rad for the Gaussian and $\thobs = 0.43\pm0.12$ rad, $\thC = 0.05 \pm 0.01$ for the power law in excellent agreement with the MCMC fits. 

% TODO: RECONSIDER, CLARIFY, OR DROP THIS
Despite the tight constraints each model gives on $\thobs$, $\thC$, and especially $\thobs/\thC$, the resulting posteriors are incompatible with each other.  The median and 68\% uncertainties in Table \ref{tab:MCMC} display only a mild tension in $\thobs$  and $\thC$ themselves, but a very large discrepancy in $\thobs/\thC$.  This is, of course, due to the very different energy profiles $E(\theta)$ in each model.  Both models can easily accommodate a rising light curve $F_\nu \sim t^{0.9}$, that is produce an effective structure parameter $\geff = 7.5$, but do so using somewhat different geometries.

%%%%%%%%%%%%%%%
%
%.  DISCUSSION
%
%%%%%%%%%%%%%%%

\section{Discussion}\label{sec:discussion}

\subsection{Inferring $\thobs$ and $E(\theta)$}

Information about the viewing angle and jet structure is clearly encoded within misaligned afterglow light curves, primarily through the temporal slope $\alpha$ during the structured phase.   This parameter alone, however, does not uniquely identify a particular structure model or observer inclination.  When trying to infer information about a particular afterglow, there is a massive degeneracy between $E(\theta)$ and $\thobs$.  

This degeneracy is both useful and unfortunate.  The presence of an extended slowly decaying or rising afterglow is a largely model independent prediction of misaligned viewing and the presence of non-trivial jet structure.  Unfortunately, this same model independence makes it very difficult to distinguish between jet structure profiles, leaving both $E(\theta)$ and $\thobs$ uncertain.

\begin{figure}
	\includegraphics[width=\columnwidth]{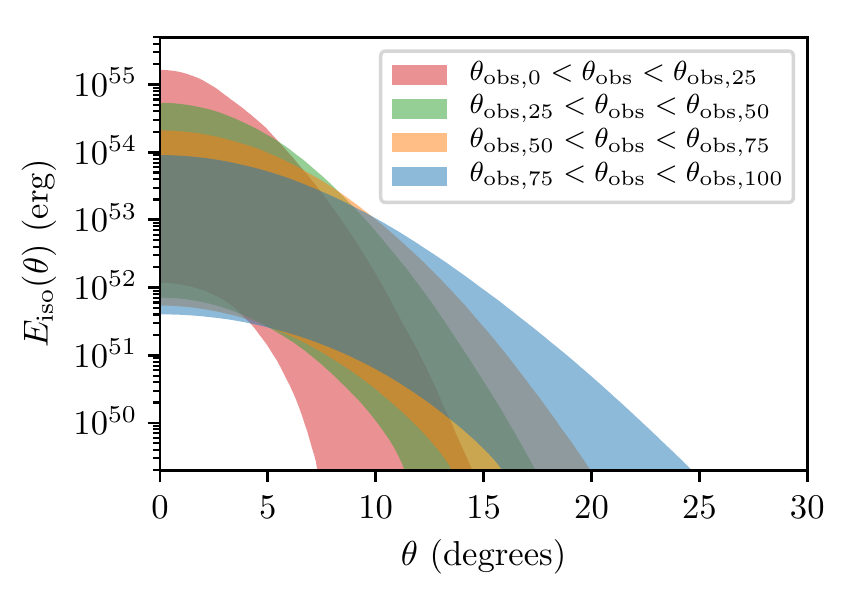}
	\caption{Inferred $E(\theta)$ for \grbbns{} in each quartile of the inferred $\thobs$ distribution, assuming Planck $H_0$ and Gaussian jet structure. Shaded bands show the symmetric 95\% confidence interval for $E(\theta)$ for $\thobs$ in the 0\%-25\% quartile (red), 25\%-50\% quartile (green), 50\%-75\% quartile (orange), 75\%-100\% quartile (blue).  The quartile boundaries are $\theta_{\mathrm{obs}, 0} = 10.9^\circ$, $\theta_{\mathrm{obs}, 25} = 19.9^\circ$, $\theta_{\mathrm{obs}, 50} = 25.0^\circ$, $\theta_{\mathrm{obs}, 75} = 29.7^\circ$, and $\theta_{\mathrm{obs}, 100} = 45.8^\circ$.  Uncertainty in $E(\theta)$ is directly correlated with uncertainty in $\thobs$: improved knowledge of one improves knowledge of the other.\label{fig:fit_E_theta}}
\end{figure}

As an example, Figure \ref{fig:fit_E_theta} shows the $95\%$ confidence intervals on $E(\theta)$ for each quartile of the $\thobs$ distribution from the Gaussian jet fit to \grbbns{}. The power law jet fit shows similar behaviour. Smaller values of $\thobs$ are correlated with more highly collimated profiles and larger energies on the jet axis, while larger values of $\thobs$ require broader profiles and less energy on the jet axis. Between the first ($\thobs < 19.9^\circ$) and fourth ($\thobs > 45.8^\circ$) quartiles the upper bound on $E_0$ lowers by more than an order of magnitude, while the bounds on energy at $\theta = 15^\circ$ shift upwards by three orders of magnitude.  This is a consequence of the tight constraint on $\thobs/\thC$ from $\alpha_{\mathrm{struct}}$ but relatively loose constraints on other parameter combinations.  At fixed $\thobs/\thC$, smaller inclinations necessarily require smaller jet opening angles and vice versa.  

Any attempt to robustly measure $\thobs$ from the afterglow light curve must be sure to allow a large variety of structure profiles $E(\theta)$ to avoid biasing the result. Conversely, any attempt to robustly measure the parameters of $E(\theta)$ must consider the full range of $\thobs$. On the other hand, any method of breaking the degeneracy and measuring $\thobs$ or $E(\theta)$ independently will also help constrain the other.

There are at least two ways of easing this degeneracy: incorporating data other than the afterglow light curve itself and looking at population-level statistics.  Utilizing posteriors from the gravitational wave signal or fitting the VLBI observations of superluminal apparent motion can greatly improve constraints on $\thobs$ \citep{Troja:2018aa, Hotokezaka:2018aa, Ghirlanda:2019aa}.  Since the afterglow presumably provides good constraints on $\geff$ (via $\alpha$), improved knowledge of $\thobs$ immediately improves knowledge of $\thC$, $b$, and any other parameters relevant to $E(\theta)$ (see for example Table 2 of \citet{Troja:2018aa}).  In some cases this may rule out a particular $E(\theta)$ entirely by pushing the relevant parameters to extreme values, for instance it is very hard to reconcile the observations of \grbbns{} with a $b=2$ power law jet.

\begin{figure}
	\includegraphics[width=\columnwidth]{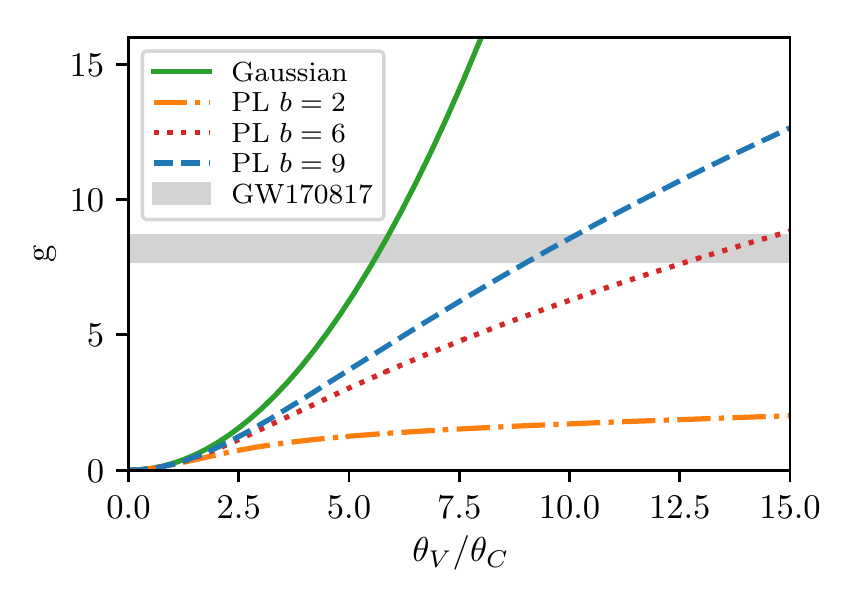}
	\caption{Structure parameter $g$ as a function of $\thobs/\thC$ for different structured jet models: Gaussian (green, solid), $b=2$ power law (orange, dash-dotted), $b=6$ power law (red, dotted), and $b=9$ power law (blue, dashed).  The inferred value $g = 8.2\pm0.5$ for \grbbns{} is shown in as the grey band. \label{fig:gPop}}
\end{figure}

In the future, it may be possible to attack the question of determining $E(\theta)$ at the population level.  Figure \ref{fig:gPop} shows the structure parameter $g$ as a function of $\thobs/\thC$ for several jet models, as well as the inferred value from \gwbns{}.  If all short GRB jets share a jet profile shape, they should trace out a single curve in this space.  As more misaligned afterglows are observed, measurements of $g$ can populate this diagram and potentially rule out models (such as potentially the $b=2$ power law).  Model-independent constraints on $\thobs/\thC$ will greatly aid this procedure.  Moreover, each structure model should predict a different observed distribution of $g$, depending on the shape of the $g(\thobs/\thC)$ curve and the detectability of each burst.  With sufficient observations of $g$ alone and understanding of the observational biases, it may be possible to infer the underlying $g(\thobs/\thC)$ curve and $E(\theta)$ itself.

\subsection{Comparison With Other \gwbns{} Analyses}

Many groups have used structured jet models to analyze the afterglow of \gwbns{}, among the most recent with comparable models are \citet{Hotokezaka:2018aa, Ghirlanda:2019aa, Lamb:2019aa, Wu:2018aa}.  Of these, \citet{Hotokezaka:2018aa, Ghirlanda:2019aa, Lamb:2019aa} use semi-analytic methods similar to the present work to construct afterglow light curves, while \citet{Wu:2018aa} utilizes analytic scaling relations and a template bank constructed from numerical simulations (following the conceptual approach introduced by \citet{van-Eerten:2012ac}).  Additionally, \citet{Hotokezaka:2018aa, Ghirlanda:2019aa} include constraints from the VLBI measurements of the \gwbns{} radio centroid's apparent superluminal motion.

\citet{Hotokezaka:2018aa} finds a very tight constraint for the viewing angle, $\thobs = 0.29 \pm 0.01$ rad and $\thobs = 0.30 \pm 0.01$ rad for their power law and Gaussian models, respectively.  The large constraining power of the apparent superluminal motion is quite evident.  They find $\thC \approx 0.05$ rad for the Gaussian jet, which lies along the $\thobs/\thC = 5.7\pm 0.2$ degeneracy from $\alpha_{\mathrm{struct}}$ for the Gaussian jet, in agreement with the closure relations.  They also find $\thC \approx 0.05$ rad for the power law jet, which is not directly comparable with our constraints due to different formulations of the power law profile.  They also find $\log_{10} E_0 / n_0 \approx 56$ for both models, in agreement with our constraints.  \citet{Ghirlanda:2019aa} uses a power law jet structure profile with the same formulation as \citet{Hotokezaka:2018aa} and sees consistent results.  The inferred parameters have somewhat larger uncertainties as they did not observe apparent superluminal motion and rely on constraints from the size of the radio centroid.

\citet{Lamb:2019aa} fits the \grbbns{} afterglow with both a Gaussian structured jet and a two-component model.  For the Gaussian they find $\thobs = 0.34 \pm 0.02$ rad and $\thC = 0.06 \pm 0.01$ rad, where we have adjusted their $\thC$ by a factor of $\sqrt{2}$ to keep in line with our definition.  Their tight bounds on both parameters are due in part to their restricted set of priors, which bound 0.3 rad < $\thobs$ < 0.4 rad.  Their $\thobs$ and $\thC$ also lie along $\thobs/\thC = 5.7 \pm 0.2$, and within their relatively narrow priors the other inferred parameters are consistent with ours in Table \ref{tab:MCMC} and Figure \ref{fig:cornerGaussian}.

\citet{Wu:2018aa} uses light curves tabulated from numerical simulations initialized with the boosted fireball model \citep{Duffell:2013aa}, parameterized by an initial bulk Lorentz factor $\gamma_B$ and specific internal energy $\eta_0$ instead of an opening angle $\thC$, although roughly $\thC \sim \gamma_B^{-1}$.  They find $\thobs = 0.47^{+0.17}_{-0.05}$ rad, somewhat larger than other works due primarily to focusing on the afterglow alone and not including GW or VLBI constraints.  They demonstrate a clear anti-correlation between $\gamma_B$ and $\thobs$ of $\gamma_B \thobs  \sim \thobs/\thC \approx 5$, and other parameters consistent with this and other studies.

It is somewhat remarkable that the disparate structured jet models used by these and other works in the literature to analyze \gwbns{} all manage to recreate the afterglow light curve with similar jet and microphysical parameters.  It also demonstrates that even with an event as well-observed as \gwbns{} it will be difficult to robustly determine the energy profile of the jet.

\subsection{O3 And Beyond}

In the coming years there will be more \gwbns{}-like events, although perhaps few with as extensive followup campaigns.  Since many of these events may be faint and distant, it will be important to leverage as much information from the afterglow light curve as possible.  As shown here, and evidenced by the diversity of models that have fit the \gwbns{} afterglow, there is a large degeneracy between viewing angle and jet structure profile.  To make robust inferences about these events, several flexible jet models must be included in the analysis.  

We hope \afterglowpy{} and similar products will aid these analyses in the future.  

\section{Summary}\label{sec:summary}

We have constructed flexible models for the electromagnetic afterglows of angularly structured relativistic jets for aligned and misaligned observers.  Through analytic approximations we can identify the basic phases of the structured afterglow light curve, determining formulae for the characteristic break time scales and closure relations.  We find the closure relations depend on a single dimensionless parameter $g$ related to the jet structure and viewing angle.  Measurements of $g$ itself are model independent, but relating it back to physical parameters of the jet is not.

We also constructed a numerical model of the afterglows of structured jets, implemented in the public \python{} package \afterglowpy{}. \afterglowpy{} computes afterglow light curves on the fly utilizing semi-analytic approximations to the jet evolution and synchrotron emission, taking into account relativistic beaming, the equal time of arrival surface, jet angular structure, trans-relativistic evolution, and jet spreading.  It is fast enough to be incorporated in MCMC parameter estimation routines, available via {\tt PyPI} and {\tt pip}, and has source code available at: \url{https://github.com/geoffryan/afterglowpy}. Fitting the \gwbns{} afterglow with multiple \afterglowpy{} models allows jet parameters to be inferred, but the degeneracy between jet structure and viewing angle makes determining a particular jet profile difficult.  Incorporating data beyond the afterglow light curve, including constraints from gravitational wave observations and detailed electromagnetic followup such as VLBI, will be key to learning as much as we can from future events.

\section{Acknowledgements} \label{sec:acknowledgements}

GR acknowledges the support from the University of Maryland through the Joint Space Science Institute Prize Postdoctoral Fellowship. LP acknowledges partial support from the agreement ASI-INAF n.2017-14-H.0. Analysis was performed on the YORP cluster administered by the Center for Theory and Computation, part of the Department of Astronomy at the University of Maryland.  GR thanks Cato Sandford for many useful discussions.

% \newpage

\appendix
\section{Derivation of the Off-Axis Jet equations}\label{app:derive1}

  The \emph{structured jet} model is a generalization of the simple top hat jet where the energy and Lorentz factor vary with the polar angle.  The light curves of structured jets display more complex behavior than top hats, which we can understand through some simple analytic relationships.
  
  Firstly, the complex behavior of a structured jet is due to relativistic beaming enhancing the jet emission at different angles as a function of time.  Once the jet becomes non-relativistic this effect is suppressed and the entire jet comes into view.  As such we will focus on the emission when the jet remains relativistic, the late-time behavior is the same as any Newtonian jet of comparable total energy.  Numerical simulations and analytic considerations have demonstrated that jet spreading does not begin in earnest until the blast wave approaches sub relativistic velocity so we will also neglect the effects of spreading and assume each sector of the jet evolves independently.  Lastly, we will assume when each sector of the blast wave is visible it is in the deceleration regime.  In this phase of evolution the blast wave Lorentz factor evolves according to:
  \begin{equation}
	\gamma(t; \theta) \propto \sqrt{\frac{E(\theta)}{n_0}}\ t^{-3/2}\ . \label{eq:app:lorentzEvolution}
\end{equation}
  In the above $\gamma$ is the Lorentz factor of the shocked fluid at an angle $\theta$ from the jet axis at lab time (in the burster frame) $t$.  The blast wave expands into a medium of constant density $n_0$ and has an angularly dependent isotropic-equivalent energy $E(\theta)$.  The forward shock is at a position $R(t; \theta)$, and moves at a speed $\beta_s = (1-\gamma_s^{-2})^{1/2}$, where $\gamma_s^2 = 2 \gamma^2$ is the Lorentz factor of the shock.  Assuming $\gamma \gg 1$ gives:
\begin{equation}
	R(t; \theta) = ct\left(1-\frac{1}{16 \gamma^2(t; \theta)}\right)\ .
\end{equation}
As in Section \ref{sec:numerical} we denote the angle between the viewer and a particular jet sector as $\psi$, its cosine as $\mu = \cos \psi$, and define $\chi = 2 \sin(\psi/2) \approx \psi$.  Photons emitted at time from a sector of the blast wave at time $t$ will be seen by the observer at $\tobs$:
\begin{equation}
	\tobs = t - \frac{\mu}{c} R(t,\theta) = (1-\mu)t + \frac{\mu}{16}\frac{t}{ \gamma^2(t,\theta)} \label{eq:app:tobs}
\end{equation}
Equation \eqref{eq:app:tobs} ignores the effect of cosmological redshift as it does not effect the closure relations.

The observed flux depends on the luminosity distance $d_L$, viewing angle $\thobs$, and rest-frame emissivity $\varepsilon'_{\nu'}$.  The Doppler factor is $\delta = \gamma^{-1} (1-\beta\mu)^{-1}$, where $\beta=(1-\gamma^{-2})^{1/2}$ is the fluid three velocity.  The observed flux can then be expressed as a volume integral, where the integrand is evaluated at the time $t$ corresponding to $\tobs$ and position ${\bf r}$.
\begin{equation}
	F_\nu(\tobs, \nuobs) = \frac{1}{4\pi d_L^2} \int d^3{\bf r} \ \delta^2 \varepsilon'_{\nu'}\ .
\end{equation}  
The blast wave emits from a region of width $\Delta R \propto \delta_s \gamma_s \gamma^{-2} R $ where $\delta_s$ is the Doppler factor associated with the shock Lorentz factor $\gamma_s$ \citep{van-Eerten:2010aa, van-Eerten:2018ab}. At a given observer time, the emission will be dominated by a region of (rest-frame) angular size $\Delta \Omega$.  The flux can then be approximated as:
\begin{equation}
	F_\nu(\tobs, \nuobs) \propto R^2 \Delta R \Delta \Omega \delta^2 \varepsilon'_{\nu'} \propto \Delta \Omega\ t^3 \gamma^{-2} \gamma_s \delta_s \delta^2 \varepsilon'_{\nu'}
\end{equation}.
The emissivity $\varepsilon'_{\nu'}$ depends on the fluid Lorentz factor, the frequency $\nuobs$, and numerous (constant) microphysical parameters.  We parameterize the dynamic dependence as $\varepsilon'_{\nu'} \propto \gamma^{s_\gamma} t^{s_t} {\nu'}^\beta = t^{s_t} \gamma^{s_\gamma}\delta^{-\beta} \nuobs^\beta$. The values of $s_\gamma$, $s_t$, and $\beta$ in synchrotron regimes are given in Table \ref{tab:specSlopes}.  This leads to a flux of:
\begin{equation}
	F_\nu \propto \Delta \Omega\ t^{3+s_t} \gamma^{-1+s_\gamma} \delta_s \delta^{2-\beta} \nuobs^\beta\ . \label{eq:app:fluxApprox}
\end{equation}

\begin{deluxetable}{CCCCC}
	\tablecaption{Dependence of rest-frame synchrotron emissivity $\epsilon'_{\nu'}$ on Lorentz factor $\gamma$ and burster-frame time $t$: $\varepsilon'_{\nu'} \propto \gamma^{s_\gamma} t^{s_t} \nu'^{\beta}$ in various spectral regimes. \label{tab:specSlopes}}
	\tablehead{\colhead{Regime}& \colhead{Label}& \colhead{$s_\gamma$} & \colhead{$s_t$} & \colhead{$\beta$}}
	\startdata
	\nu'<\nu'_m<\nu'_c		& D	& 1 & 0 & 1/3 \\
	\nu' < \nu'_c < \nu'_m 	& E	& 7/3 & 2/3 & 1/3 \\
	\nu'_c < \nu' < \nu'_m 	& F	& 3/2 & -1 & -1/2 \\
	\nu'_m<\nu'<\nu'_c     	& G	& (3p+1)/2 & 0 & (1-p)/2 \\
	\nu'_m, \nu'_c < \nu' 		& H 	& 3p/2 & -1 & -p/2 \\ 
	\enddata
\end{deluxetable}
The Doppler factor depends on the fluid Lorentz factor and whether the material is on-axis. The on/off axis boundary occurs at $\gamma^{-1} \approx \chi$ (ie. $\beta \approx \mu$).  One can write:
\begin{equation}
	\delta = \begin{cases}
				2 \gamma^{-1} \chi^{-2},  & \text{if } \gamma \gg 1 \text{ and } \chi \gg \gamma^{-1} \ \ \text{(off-axis)} \\
				2 \gamma, & \text{if } \gamma \gg 1 \text{ and } \chi \ll \gamma^{-1} \ \  \text{(on-axis)} \\
				1  & \text{if } \beta \ll 1\ \  \text{(non-relativistic)}
		     \end{cases}
\end{equation}
The observer time can be similarly simplified in both limits:
\begin{equation}
	\tobs = \left \{ \begin{matrix}
				\frac{1}{2} \chi^2 t,  & \text{if } \chi \gg \gamma^{-1}\ \ \text{(off-axis)} \\
				\frac{1}{16} \gamma^{-2} t, & \text{if } \chi \ll \gamma^{-1} \ \  \text{(on-axis)} \end{matrix} \right . \ .
\end{equation}
As can the flux:
\begin{equation}
	F_\nu \propto \left \{ \begin{matrix}
				\Delta \Omega t^{3+s_2} \gamma^{-4+s_1+\beta} \chi^{-6+2\beta}\nuobs^\beta,  & \text{(off-axis)} \\
				\Delta \Omega t^{3+s_2} \gamma^{2+s_1-\beta} \nuobs^\beta, & \text{(on-axis)} \end{matrix} \right . \ .
\end{equation}
The behavior of the Doppler factor informs the scaling of $\Delta \Omega$.  If any part of the jet is on-axis, its emission is enhanced by $\sim \gamma^2$ over the off-axis material and will dominate.  An observer for whom the entire jet is off-axis must be situated at some large $\thobs$, outside the outermost jet material.  At early times the entire jet will be beamed off-axis, with emission from the near edge (with the smallest $\psi$ and presumably $\gamma$) contributing most to the emission.  The absence of any particular angular scale in this regime indicates $\Delta \Omega$ will be roughly constant.  For an on-axis observer $\Delta \Omega \sim \sin^2 \psi_{\rm{max}} \propto \gamma^{-2}$ until the entire jet is on-axis, at which point $\Delta \Omega$ is again constant.  Hence:
\begin{equation}
	F_\nu \propto \left \{ \begin{matrix}
				t^{3+s_2} \gamma^{-4+s_1+\beta} \chi^{-6+2\beta}\nuobs^\beta,  & \text{(off-axis)} \\
				t^{3+s_2} \gamma^{s_1-\beta} \nuobs^\beta, &  \text{(on-axis, pre jet break)} \\
				t^{3+s_2} \gamma^{2+s_1-\beta} \nuobs^\beta, & \text{(on-axis, post jet break)} \end{matrix} \right . \ .
\end{equation}
Finally, for off-axis emission $t \propto \tobs$ and hence $\gamma \propto \tobs^{-3/2}$.  For on-axis observers $\tobs \propto \gamma^{-2} t \propto t^4$, hence $t\propto \tobs^{1/4}$ and $\gamma \propto \tobs^{-3/8}$. Giving finally:
\begin{equation}
	F_\nu \propto \left \{ \begin{matrix}
				\tobs^{9+s_2 -3(s_1+\beta)/2} \chi^{-6+2\beta}\nuobs^\beta,  & \text{(off-axis)} \\
				\tobs^{(3+s_2)/4+3(-s_1+\beta)/8} \nuobs^\beta, &\text{(on-axis, pre jet break)} \\
				\tobs^{s_2/4 + 3(-s_1+\beta)/8} \nuobs^\beta, & \text{(on-axis, post jet break)} \end{matrix} \right . \ .
\end{equation}
These formulae capture the standard behavior of top-hat jets, as well as any jet that is \emph{fully} on-axis or off-axis.  What they fail to (easily) demonstrate is the behavior of a structured jet which transitions continuously from one state to the other over the course of observation.

\section{Derivation of the Structured Jet equations}\label{app:derive2}

A jet with a non-trivial angular distribution of energy can exhibit qualitatively different behaviour than a simple top hat, particularly when observed at a significant viewing angle.  While the initial off-axis and final on-axis post jet-break evolutions are identical, these are separated by a transition phase where the sector dominating the emission scans over the jet surface.  This transition phase begins at the end of the off-axis phase: when a sector of the jet first decelerates to include the observer in its beaming cone.  This will necessarily be from the wings/edge of the jet, the material with the lowest Lorentz factor and smallest angle $\psi$ to the observer.  As the blastwave decelerates more energetic material from nearer the core will come into view and the high latitude emission will dim.  Finally the core of the jet ($\theta = 0$ or $\psi = \thobs$) decelerates and becomes visible to the observer.  At this point the entire jet is on-axis and evolution continues as in the post jet-break phase.

At each moment during the structure phase the emission is dominated by material that just came on-axis, where $\gamma^{-1} = \chi \equiv 2\sin (\psi/2) \approx \psi$.  To find the overall behaviour we first determine $\tobs(\psi)$ and $F_\nu(\psi)$ for material whose emission is peaking (coming on-axis).  Since the structure phase occurs when motion is still relativistic, we can assume $\gamma \gg 1$ and hence $\chi, \sin \psi \ll 1$.  In this approximation:
\begin{eqnarray}
	\mu = 1 - \frac{1}{2}\chi^2 \ , \label{eq:app:muPsi}\\
	\delta_s \approx \delta \approx \gamma = \chi^{-1} \ . \label{eq:app:deltaPsi}
\end{eqnarray}
The material dominating the emission is in the plane between the observer and the jet axis, denoted by $\phi = 0$. Along this line we have $\psi = \thobs - \theta$, and can use $\psi$ or $\theta$ interchangeably to denote latitude.  From Equation \eqref{eq:app:lorentzEvolution} we have $t\propto E(\theta)^{1/3} \gamma^{-2/3}$.  Using Equation \eqref{eq:app:tobs} and \eqref{eq:app:muPsi} we find that material at $\psi$ will come on-axis at observer time:
\begin{equation}
	\tobs(\psi) = \frac{9}{16} \chi^2 t \propto E(\theta)^{1/3} \chi^{8/3}\ . \label{eq:app:tobsPsi}
\end{equation}
The peak flux from material at $\psi$ can be determined from Equation \eqref{eq:app:fluxApprox}.  Using Equation \eqref{eq:app:deltaPsi} and taking $\Delta \Omega \propto \gamma^{-\som}$ gives an observed flux:
\begin{equation}
	F_\nu(\psi) \propto t^{3+s_t} \gamma^{2+s_\gamma-\som-\beta} \nuobs^\beta \propto E(\theta)^{1+s_t/3} \chi^{-s_\gamma + 2 s_t/3 +\som+\beta} \nuobs^\beta  \ . \label{eq:app:FnuPsi}
\end{equation}
	Equations \eqref{eq:app:tobsPsi} and \eqref{eq:app:FnuPsi} describe the evolution of the flux in the structure phase in terms of the parameter $\psi$, which varies from $\mathrm{min} (0, \thobs-\thW)$ to $\thobs$.  In principle one would like to invert Equation \eqref{eq:app:tobsPsi} and substitute into Equation \eqref{eq:app:FnuPsi} to obtain $F_\nu(\tobs)$ itself.  Unfortunately, in general this is impossible to do in closed form because $E(\theta)$ is non-trivial. 
	
	We can obtain the temporal power law slope of the light curve by differentiating both Equations \eqref{eq:app:tobsPsi} and \eqref{eq:app:FnuPsi} with respect to $\psi$.  Noting that $dE/d\psi = -dE/d\theta$ and $d\chi/d\psi = \cos (\psi/2)$ we obtain for the individual derivatives:
\begin{eqnarray}
	\frac{d \log \tobs}{d \psi} = \frac{4}{3} \cot (\psi/2) - \frac{1}{3} \frac{d \log E}{d \theta}\ , \\
	\frac{d \log F_\nu}{d \psi} = \left(-s_\gamma + \frac{2}{3} s_t +\som+\beta\right)\frac{1}{2}\cot(\psi/2) - \left(1+\frac{1}{3}s_t\right) \frac{d \log E}{d \theta}\ .
\end{eqnarray}
Taking the ratio and simplifying gives:
\begin{eqnarray}
	\frac{d \log F_\nu}{d \log \tobs}(\psi) = \frac{3 \beta - 3s_\gamma + 2s_t+3\som + (3+s_t)g(\psi)}{ 8+g(\psi)}  \label{eq:app:closure} \\
	g(\psi) \equiv -2\tan(\psi/2) \frac{d \log E}{d \theta}\ .
\end{eqnarray}
The parameter $g$ is directly measurable from the light curve, given the spectral information which fixes $\beta$, $s_\gamma$, and $s_t$.

Equation \eqref{eq:app:closure} is a generic expression for the temporal slope $\alpha$ for synchrotron emission from and relativistic, non-spreading, decelerating blast wave.  In particular, if $g = 0$ and $\som = 2$ one recovers the standard pre-jet break slopes \citep{Granot:2002aa}.  If $g=0$ and $s=0$ one recovers the standard beaming-effect jet break where the post-break slope is reduced by $-3/4$ from the the pre-break slope.  Misaligned viewing of a structured jet corresponds to $\som = 1$, $g$ free.

The full flux scaling equations also require an updated energy and circumburst density scaling. We can obtain the scalings for energy and density from dimensional analysis, following \citet{van-Eerten:2012ac} (specifically, by making use of the fact that the overall flux scalings for the different spectral regimes should obey those presented in table 1 of that paper).

Introducing the microphysical parameters $p$, $\epse$, and $\epsB$ as in Section \ref{sec:numerical} we can write scalings for the characteristic spectral quantities in the observers frame, where we only neglect constant factors:
\begin{align}
	\nu_m (\psi) &\propto \left(1+z\right)^{-1}\left(\frac{p-2}{p-1}\right)^2 n_0^{1/2} \epse^2 \epsB^{1/2}  \chi^{-4} \label{eq:app:numPsi} \\
	\nu_c (\psi)&\propto  \left(1+z\right)^{-1} n_0^{-5/6} \epsB^{-3/2} E(\theta)^{-2/3} \chi^{-4/3} \label{eq:app:nucPsi} \\
	F_P(\psi)&\propto \left(1+z\right) \left(p-1\right)  d_L^{-2} n_0^{1/2} \epsB^{1/2} E(\theta)  \chi^{-2+\som} \label{eq:app:FPPsi}
\end{align}

Their derivatives with respect to $\psi$ are:
\begin{align}
	\frac{d \log \nu_m}{d\psi} &= -2 \cot(\psi/2)\ , \\
	\frac{d \log \nu_c}{d\psi} &= \frac{2}{3} \cot(\psi/2) - \frac{2}{3} \frac{d \log E}{d \theta}\ , \\
	\frac{d \log F_P}{d\psi} &= \left(-2 + \som\right)\frac{1}{2} \cot(\psi/2) -  \frac{d \log E}{d \theta}\ .
\end{align}
Leading to observed temporal slopes of:
\begin{align}
	\frac{d \log \nu_m}{d\log \tobs} &= -\frac{12}{8+g(\psi)}\ , \\
	\frac{d \log \nu_c}{d\log\tobs} &= 2\frac{2 + g(\psi)}{8+g(\psi)}\ , \\
	\frac{d \log F_P}{d\log \tobs} &= 3\frac{-2 + \som +  g(\psi)}{8+g(\psi)}\ .
\end{align}

We can obtain full observer frame flux scaling equations in each of the synchrotron spectral regimes, following \citet{van-Eerten:2012ac}. Although the scalings of $p$, $\epse$, $\epsB$, and $d_L$ remain unchanged, $n_0$, $E_0$, and $z$ are significantly altered due to their role in normalizing the observer time $\tobs$. In particular, by dimensional analysis $\chi$ must be proportional to a power of $(1+z)^{-1} E_0^{-1/3} n_0^{1/3} \tobs$.  We can use this to write the characteristic spectral quantities semi-explicitly in terms of the observer time $\tobs$:
\begin{align}
	\nu_m & \propto  (1+z)^{(4-g)/(8+g)}  \left(\frac{p-2}{p-1}\right)^2  \epse^2 \epsB^{1/2}E_0^{4/(8+g)} n_0^{g/(16+2g)} \tobs^{-12/(8+g)} \\
	\nu_c & \propto  (1+z)^{(-4+g)/(8+g)} \epsB^{-3/2}   E_0^{-4/(8+g)} n_0^{-(16+3g)/(16+2g)} \tobs^{2(2+g)/(8+g)} \\
	F_{\mathrm{Peak}} &\propto  (1+z)^{(14-2g-3\som)/(8+g)}  \epsB^{1/2} E_0^{(10-\som)/(8+g)} n_0^{(4+3g+2\som)/(16+2g)} t_{obs}^{3(-2+\som+g)/(8+g)}
\end{align}
Formally the above relations are not explicit, as $g$ depends on $\psi$ which depends on $\tobs$ in a non-trivial way.  However, as demonstrated in Section \ref{sec:structuredJets} in many cases $g$ can be viewed as effectively constant.  Using $\geff$ (ie. Equation \eqref{eq:geff}, for instance) one can treat the above formulae as explicit in $\tobs$.

One can then arrive at the full flux relations, where we have only dropped constant numerical factors and a rational function of $p$:
\begin{align}
F_D & \propto  (1+z)^{(38-5g-9\som)/(24+3g)} \epsilon_e^{-2/3} \epsilon_B^{1/3} n_0^{(6+4g+3\som)/(24+3g)} E_0^{(26-3\som)/(24+8g)} t_{obs}^{(-2+3\som+3g)/(8+g)} \nu^{1/3}  \\
F_E & \propto  (1+z)^{(46-7g-9\som)/(24+3g)} \epsilon_B^{1} n_0^{(14+6g+3\som)/(24+3g)} E_0^{(34-3\som)/(24+3g)} t_{obs}^{(-14+9\som+11g)/(24+3g)} \nu^{1/3}  \\
F_F & \propto  (1+z)^{(24-3g-6\som)/(16+2g)} \epsilon_B^{-1/4} n_0^{(-8+3g+4\som)/(32+4g)} E_0^{(8-\som)/(8+g)} t_{obs}^{(-8+3\som+2g)/(8+g)} \nu^{-1/2}  \\
F_G & \propto  (1+z)^{(-(p+3)g+4p-6\som+24)/(16+2g)} \epsilon_e^{p-1} \epsilon_B^{(1+p)/4} n_0^{(gp + 5 g + 4\som + 8)/(32+4g)} E_0^{(8+2p-\som)/(8+g)} \nonumber \\
 	& \qquad \times \tobs^{3(-2p+\som+g)/(8+g)} \nu^{(1-p)/2}  \\
F_H & \propto  (1+z)^{(-(p+2)g+4p-6\som+20)/(16+2g)} \epsilon_e^{p-1} \epsilon_B^{(p-2)/4} n_0^{(gp+2g+4\som-8)/(32+4g)} E_0^{(6+2p-\som)/(8+g)} \nonumber \\
	&\qquad \times \tobs^{(-6p-2+3\som+2g)/(8+g)} \nu^{-p/2} 
\end{align}
Again, although formally $g = g(\psi)$ is time variable, it can often be regarded as constant in a particular phase of the light curve.

\section{GW170817A - Parameter Estimation Posteriors}\label{sec:corners}

\begin{figure*}
	\includegraphics[width=\textwidth]{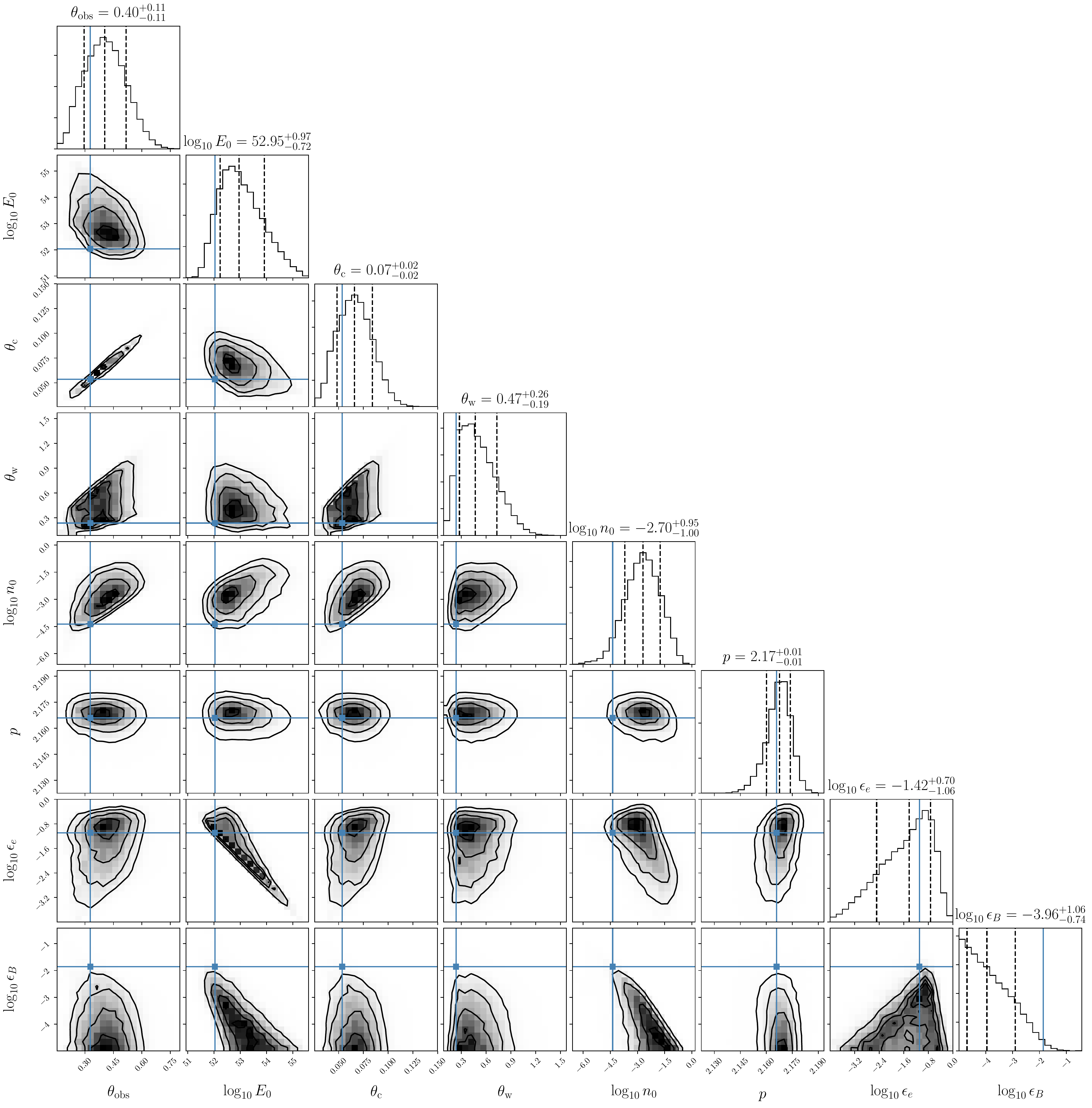}
	\caption{Views of the posterior parameter distribution for a Gaussian jet fit to the \gwbns{} afterglow.  The diagonal contains one-dimensional marginalized posteriors for each fit parameter, while the off-diagonal plots contain two-dimensional maps of the posterior marginalized over all but the two corresponding parameters. Dashed lines and labels along the diagonal give the median value and symmetric 68\% uncertainties (the 16\% and 84\% quantiles) for each parameter's marginalized distribution.  Blue dashed line shows the location of the sample with maximum posterior probability. \label{fig:cornerGaussian}}
\end{figure*}

\begin{figure*}
	\includegraphics[width=\textwidth]{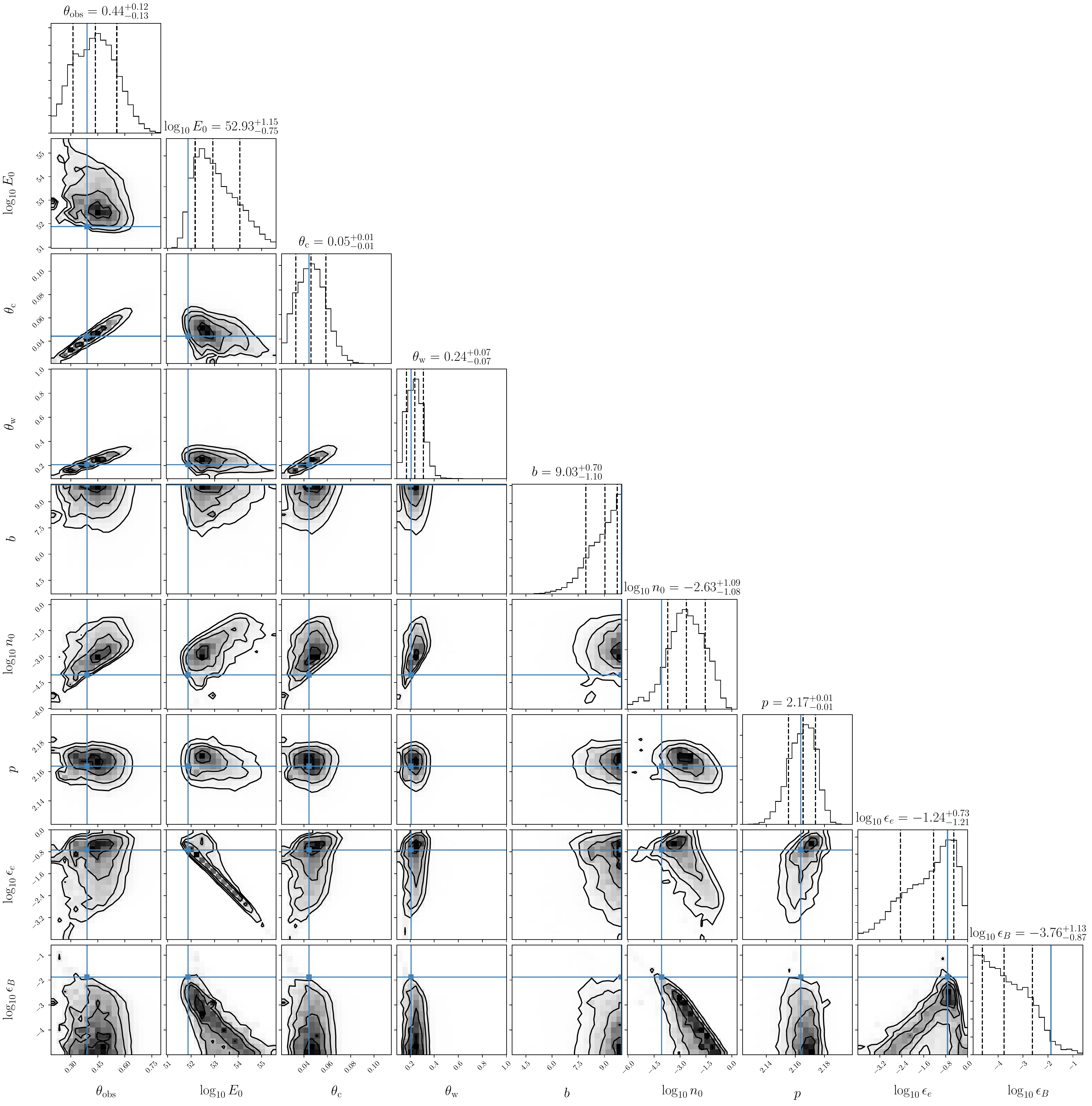}
	\caption{As Figure \ref{fig:cornerGaussian} for the power law jet model. \label{fig:cornerPowerlaw}}
\end{figure*}

The marginalized posterior limits on the afterglow parameters presented in Table \ref{tab:MCMC} are a useful summary but do not show how the inferred parameter values correlate with each other.  Figures \ref{fig:cornerGaussian} and \ref{fig:cornerPowerlaw} show all pairwise correlations between parameters in our fits to \grbbns{} for the Gaussian and power law jet models respectively, as well as the marginalized distributions for each parameter.

\bibliography{structuredJets_sources}

\begin{thebibliography}{}
\expandafter\ifx\csname natexlab\endcsname\relax\def\natexlab#1{#1}\fi
\providecommand{\url}[1]{\href{#1}{#1}}
\providecommand{\dodoi}[1]{doi:~\href{http://doi.org/#1}{\nolinkurl{#1}}}
\providecommand{\doeprint}[1]{\href{http://ascl.net/#1}{\nolinkurl{http://ascl.net/#1}}}
\providecommand{\doarXiv}[1]{\href{https://arxiv.org/abs/#1}{\nolinkurl{https://arxiv.org/abs/#1}}}

\bibitem[{{Abbott} {et~al.}(2017{\natexlab{a}}){Abbott}, {Abbott}, {Abbott},
  {Acernese}, {Ackley}, {Adams}, {Adams}, {Addesso}, {Adhikari}, {Adya}, \&
  et~al.}]{Abbott:2017ab}
{Abbott}, B.~P., {Abbott}, R., {Abbott}, T.~D., {et~al.} 2017{\natexlab{a}},
  \prl, 119, 161101, \dodoi{10.1103/PhysRevLett.119.161101}

\bibitem[{{Abbott} {et~al.}(2017{\natexlab{b}}){Abbott}, {Abbott}, {Abbott},
  {Acernese}, {Ackley}, {Adams}, {Adams}, {Addesso}, {Adhikari}, {Adya}, \&
  et~al.}]{Abbott:2017ac}
---. 2017{\natexlab{b}}, \apjl, 848, L12, \dodoi{10.3847/2041-8213/aa91c9}

\bibitem[{{Abbott} {et~al.}(2017{\natexlab{c}}){Abbott}, {Abbott}, {Abbott},
  {Acernese}, {Ackley}, {Adams}, {Adams}, {Addesso}, {Adhikari}, {Adya}, \&
  et~al.}]{Abbott:2017aa}
---. 2017{\natexlab{c}}, \nat, 551, 85, \dodoi{10.1038/nature24471}

\bibitem[{{Alexander} {et~al.}(2018){Alexander}, {Margutti}, {Blanchard},
  {Fong}, {Berger}, {Hajela}, {Eftekhari}, {Chornock}, {Cowperthwaite},
  {Giannios}, {Guidorzi}, {Kathirgamaraju}, {MacFadyen}, {Metzger}, {Nicholl},
  {Sironi}, {Villar}, {Williams}, {Xie}, \& {Zrake}}]{Alexander:2018aa}
{Alexander}, K.~D., {Margutti}, R., {Blanchard}, P.~K., {et~al.} 2018, \apjl,
  863, L18, \dodoi{10.3847/2041-8213/aad637}

\bibitem[{{Aloy} {et~al.}(2005){Aloy}, {Janka}, \& {M{\"u}ller}}]{Aloy:2005aa}
{Aloy}, M.~A., {Janka}, H.-T., \& {M{\"u}ller}, E. 2005, \aap, 436, 273,
  \dodoi{10.1051/0004-6361:20041865}

\bibitem[{{Blandford} \& {McKee}(1976)}]{Blandford:1976aa}
{Blandford}, R.~D., \& {McKee}, C.~F. 1976, Physics of Fluids, 19, 1130,
  \dodoi{10.1063/1.861619}

\bibitem[{{Corsi} \& {Piro}(2006)}]{Corsi:2006aa}
{Corsi}, A., \& {Piro}, L. 2006, \aap, 458, 741,
  \dodoi{10.1051/0004-6361:20065280}

\bibitem[{{Dalal} {et~al.}(2002){Dalal}, {Griest}, \& {Pruet}}]{Dalal:2002aa}
{Dalal}, N., {Griest}, K., \& {Pruet}, J. 2002, \apj, 564, 209,
  \dodoi{10.1086/324142}

\bibitem[{{D'Alessio} {et~al.}(2006){D'Alessio}, {Piro}, \&
  {Rossi}}]{DAlessio:2006aa}
{D'Alessio}, V., {Piro}, L., \& {Rossi}, E.~M. 2006, \aap, 460, 653,
  \dodoi{10.1051/0004-6361:20054501}

\bibitem[{{D'Avanzo} {et~al.}(2018){D'Avanzo}, {Campana}, {Salafia},
  {Ghirlanda}, {Ghisellini}, {Melandri}, {Bernardini}, {Branchesi},
  {Chassande-Mottin}, {Covino}, {D'Elia}, {Nava}, {Salvaterra}, {Tagliaferri},
  \& {Vergani}}]{DAvanzo:2018aa}
{D'Avanzo}, P., {Campana}, S., {Salafia}, O.~S., {et~al.} 2018, \aap, 613, L1,
  \dodoi{10.1051/0004-6361/201832664}

\bibitem[{{Duffell} \& {Laskar}(2018)}]{Duffell:2018aa}
{Duffell}, P.~C., \& {Laskar}, T. 2018, \apj, 865, 94,
  \dodoi{10.3847/1538-4357/aadb9c}

\bibitem[{{Duffell} \& {MacFadyen}(2013)}]{Duffell:2013aa}
{Duffell}, P.~C., \& {MacFadyen}, A.~I. 2013, \apjl, 776, L9,
  \dodoi{10.1088/2041-8205/776/1/L9}

\bibitem[{{Fong} {et~al.}(2015){Fong}, {Berger}, {Margutti}, \&
  {Zauderer}}]{Fong:2015aa}
{Fong}, W., {Berger}, E., {Margutti}, R., \& {Zauderer}, B.~A. 2015, \apj, 815,
  102, \dodoi{10.1088/0004-637X/815/2/102}

\bibitem[{{Fong} {et~al.}(2019){Fong}, {Blanchard}, {Alexander}, {Strader},
  {Margutti}, {Hajela}, {Villar}, {Wu}, {Ye}, {Berger}, {Chornock},
  {Coppejans}, {Cowperthwaite}, {Eftekhari}, {Giannios}, {Guidorzi},
  {Kathirgamaraju}, {Laskar}, {MacFadyen}, {Metzger}, {Nicholl}, {Paterson},
  {Terreran}, {Sand}, {Sironi}, {Williams}, {Xie}, \& {Zrake}}]{Fong:2019aa}
{Fong}, W.-f., {Blanchard}, P.~K., {Alexander}, K.~D., {et~al.} 2019, arXiv
  e-prints, arXiv:1908.08046.
\newblock \doarXiv{1908.08046}

\bibitem[{{Foreman-Mackey} {et~al.}(2013){Foreman-Mackey}, {Hogg}, {Lang}, \&
  {Goodman}}]{Foreman-Mackey:2013aa}
{Foreman-Mackey}, D., {Hogg}, D.~W., {Lang}, D., \& {Goodman}, J. 2013, \pasp,
  125, 306, \dodoi{10.1086/670067}

\bibitem[{{Ghirlanda} {et~al.}(2019){Ghirlanda}, {Salafia}, {Paragi},
  {Giroletti}, {Yang}, {Marcote}, {Blanchard}, {Agudo}, {An}, {Bernardini},
  {Beswick}, {Branchesi}, {Campana}, {Casadio}, {Chassande-Mottin}, {Colpi},
  {Covino}, {D'Avanzo}, {D'Elia}, {Frey}, {Gawronski}, {Ghisellini}, {Gurvits},
  {Jonker}, {van Langevelde}, {Melandri}, {Moldon}, {Nava}, {Perego},
  {Perez-Torres}, {Reynolds}, {Salvaterra}, {Tagliaferri}, {Venturi},
  {Vergani}, \& {Zhang}}]{Ghirlanda:2019aa}
{Ghirlanda}, G., {Salafia}, O.~S., {Paragi}, Z., {et~al.} 2019, Science, 363,
  968, \dodoi{10.1126/science.aau8815}

\bibitem[{{Granot} \& {Kumar}(2003)}]{Granot:2003aa}
{Granot}, J., \& {Kumar}, P. 2003, \apj, 591, 1086, \dodoi{10.1086/375489}

\bibitem[{{Granot} {et~al.}(2002){Granot}, {Panaitescu}, {Kumar}, \&
  {Woosley}}]{Granot:2002ab}
{Granot}, J., {Panaitescu}, A., {Kumar}, P., \& {Woosley}, S.~E. 2002, \apjl,
  570, L61, \dodoi{10.1086/340991}

\bibitem[{{Granot} \& {Piran}(2012)}]{Granot:2012aa}
{Granot}, J., \& {Piran}, T. 2012, \mnras, 421, 570,
  \dodoi{10.1111/j.1365-2966.2011.20335.x}

\bibitem[{{Granot} \& {Sari}(2002)}]{Granot:2002aa}
{Granot}, J., \& {Sari}, R. 2002, \apj, 568, 820, \dodoi{10.1086/338966}

\bibitem[{{Haggard} {et~al.}(2017){Haggard}, {Nynka}, {Ruan}, {Kalogera},
  {Cenko}, {Evans}, \& {Kennea}}]{Haggard:2017aa}
{Haggard}, D., {Nynka}, M., {Ruan}, J.~J., {et~al.} 2017, \apjl, 848, L25,
  \dodoi{10.3847/2041-8213/aa8ede}

\bibitem[{{Hallinan} {et~al.}(2017){Hallinan}, {Corsi}, {Mooley}, {Hotokezaka},
  {Nakar}, {Kasliwal}, {Kaplan}, {Frail}, {Myers}, {Murphy}, {De}, {Dobie},
  {Allison}, {Bannister}, {Bhalerao}, {Chandra}, {Clarke}, {Giacintucci}, {Ho},
  {Horesh}, {Kassim}, {Kulkarni}, {Lenc}, {Lockman}, {Lynch}, {Nichols},
  {Nissanke}, {Palliyaguru}, {Peters}, {Piran}, {Rana}, {Sadler}, \&
  {Singer}}]{Hallinan:2017aa}
{Hallinan}, G., {Corsi}, A., {Mooley}, K.~P., {et~al.} 2017, Science, 358,
  1579, \dodoi{10.1126/science.aap9855}

\bibitem[{{Hotokezaka} {et~al.}(2018){Hotokezaka}, {Nakar}, {Gottlieb},
  {Nissanke}, {Masuda}, {Hallinan}, {Mooley}, \& {Deller}}]{Hotokezaka:2018aa}
{Hotokezaka}, K., {Nakar}, E., {Gottlieb}, O., {et~al.} 2018, arXiv e-prints.
\newblock \doarXiv{1806.10596}

\bibitem[{{Hotokezaka} \& {Piran}(2015)}]{Hotokezaka:2015aa}
{Hotokezaka}, K., \& {Piran}, T. 2015, \mnras, 450, 1430,
  \dodoi{10.1093/mnras/stv620}

\bibitem[{{Ioka} \& {Nakamura}(2001)}]{Ioka:2001aa}
{Ioka}, K., \& {Nakamura}, T. 2001, \apjl, 554, L163, \dodoi{10.1086/321717}

\bibitem[{{Kumar} \& {Granot}(2003)}]{Kumar:2003aa}
{Kumar}, P., \& {Granot}, J. 2003, \apj, 591, 1075, \dodoi{10.1086/375186}

\bibitem[{{Lamb} \& {Kobayashi}(2017)}]{Lamb:2017aa}
{Lamb}, G.~P., \& {Kobayashi}, S. 2017, \mnras, 472, 4953,
  \dodoi{10.1093/mnras/stx2345}

\bibitem[{{Lamb} {et~al.}(2019){Lamb}, {Lyman}, {Levan}, {Tanvir}, {Kangas},
  {Fruchter}, {Gompertz}, {Hjorth}, {Mandel}, {Oates}, {Steeghs}, \&
  {Wiersema}}]{Lamb:2019aa}
{Lamb}, G.~P., {Lyman}, J.~D., {Levan}, A.~J., {et~al.} 2019, \apjl, 870, L15,
  \dodoi{10.3847/2041-8213/aaf96b}

\bibitem[{{Lazzati} {et~al.}(2017){Lazzati}, {L{\'o}pez-C{\'a}mara},
  {Cantiello}, {Morsony}, {Perna}, \& {Workman}}]{Lazzati:2017aa}
{Lazzati}, D., {L{\'o}pez-C{\'a}mara}, D., {Cantiello}, M., {et~al.} 2017,
  \apjl, 848, L6, \dodoi{10.3847/2041-8213/aa8f3d}

\bibitem[{{Lipunov} {et~al.}(2001){Lipunov}, {Postnov}, \&
  {Prokhorov}}]{Lipunov:2001aa}
{Lipunov}, V.~M., {Postnov}, K.~A., \& {Prokhorov}, M.~E. 2001, Astronomy
  Reports, 45, 236, \dodoi{10.1134/1.1353364}

\bibitem[{{Lyman} {et~al.}(2018){Lyman}, {Lamb}, {Levan}, {Mandel}, {Tanvir},
  {Kobayashi}, {Gompertz}, {Hjorth}, {Fruchter}, {Kangas}, {Steeghs}, {Steele},
  {Cano}, {Copperwheat}, {Evans}, {Fynbo}, {Gall}, {Im}, {Izzo}, {Jakobsson},
  {Milvang-Jensen}, {O'Brien}, {Osborne}, {Palazzi}, {Perley}, {Pian},
  {Rosswog}, {Rowlinson}, {Schulze}, {Stanway}, {Sutton}, {Th{\"o}ne}, {de
  Ugarte Postigo}, {Watson}, {Wiersema}, \& {Wijers}}]{Lyman:2018aa}
{Lyman}, J.~D., {Lamb}, G.~P., {Levan}, A.~J., {et~al.} 2018, Nature Astronomy,
  2, 751, \dodoi{10.1038/s41550-018-0511-3}

\bibitem[{{Margutti} {et~al.}(2018){Margutti}, {Alexander}, {Xie}, {Sironi},
  {Metzger}, {Kathirgamaraju}, {Fong}, {Blanchard}, {Berger}, {MacFadyen},
  {Giannios}, {Guidorzi}, {Hajela}, {Chornock}, {Cowperthwaite}, {Eftekhari},
  {Nicholl}, {Villar}, {Williams}, \& {Zrake}}]{Margutti:2018aa}
{Margutti}, R., {Alexander}, K.~D., {Xie}, X., {et~al.} 2018, \apjl, 856, L18,
  \dodoi{10.3847/2041-8213/aab2ad}

\bibitem[{{M{\'e}sz{\'a}ros} {et~al.}(1998){M{\'e}sz{\'a}ros}, {Rees}, \&
  {Wijers}}]{Meszaros:1998aa}
{M{\'e}sz{\'a}ros}, P., {Rees}, M.~J., \& {Wijers}, R.~A.~M.~J. 1998, \apj,
  499, 301, \dodoi{10.1086/305635}

\bibitem[{{Mignone} {et~al.}(2005){Mignone}, {Plewa}, \&
  {Bodo}}]{Mignone:2005aa}
{Mignone}, A., {Plewa}, T., \& {Bodo}, G. 2005, The Astrophysical Journal
  Supplement Series, 160, 199, \dodoi{10.1086/430905}

\bibitem[{{Mizuta} \& {Aloy}(2009)}]{Mizuta:2009aa}
{Mizuta}, A., \& {Aloy}, M.~A. 2009, \apj, 699, 1261,
  \dodoi{10.1088/0004-637X/699/2/1261}

\bibitem[{{Mooley} {et~al.}(2018{\natexlab{a}}){Mooley}, {Nakar}, {Hotokezaka},
  {Hallinan}, {Corsi}, {Frail}, {Horesh}, {Murphy}, {Lenc}, {Kaplan}, {de},
  {Dobie}, {Chandra}, {Deller}, {Gottlieb}, {Kasliwal}, {Kulkarni}, {Myers},
  {Nissanke}, {Piran}, {Lynch}, {Bhalerao}, {Bourke}, {Bannister}, \&
  {Singer}}]{Mooley:2018aa}
{Mooley}, K.~P., {Nakar}, E., {Hotokezaka}, K., {et~al.} 2018{\natexlab{a}},
  \nat, 554, 207, \dodoi{10.1038/nature25452}

\bibitem[{{Mooley} {et~al.}(2018{\natexlab{b}}){Mooley}, {Deller}, {Gottlieb},
  {Nakar}, {Hallinan}, {Bourke}, {Frail}, {Horesh}, {Corsi}, \&
  {Hotokezaka}}]{Mooley:2018ab}
{Mooley}, K.~P., {Deller}, A.~T., {Gottlieb}, O., {et~al.} 2018{\natexlab{b}},
  \nat, 561, 355, \dodoi{10.1038/s41586-018-0486-3}

\bibitem[{{Nakar} {et~al.}(2018){Nakar}, {Gottlieb}, {Piran}, {Kasliwal}, \&
  {Hallinan}}]{Nakar:2018aa}
{Nakar}, E., {Gottlieb}, O., {Piran}, T., {Kasliwal}, M.~M., \& {Hallinan}, G.
  2018, The Astrophysical Journal, 867, 18, \dodoi{10.3847/1538-4357/aae205}

\bibitem[{{Nakar} {et~al.}(2004){Nakar}, {Granot}, \& {Guetta}}]{Nakar:2004aa}
{Nakar}, E., {Granot}, J., \& {Guetta}, D. 2004, \apjl, 606, L37,
  \dodoi{10.1086/421107}

\bibitem[{{Nakar} \& {Piran}(2011)}]{Nakar:2011aa}
{Nakar}, E., \& {Piran}, T. 2011, \nat, 478, 82, \dodoi{10.1038/nature10365}

\bibitem[{{Nava} {et~al.}(2013){Nava}, {Sironi}, {Ghisellini}, {Celotti}, \&
  {Ghirlanda}}]{Nava:2013aa}
{Nava}, L., {Sironi}, L., {Ghisellini}, G., {Celotti}, A., \& {Ghirlanda}, G.
  2013, \mnras, 433, 2107, \dodoi{10.1093/mnras/stt872}

\bibitem[{{Nousek} {et~al.}(2006){Nousek}, {Kouveliotou}, {Grupe}, {Page},
  {Granot}, {Ramirez-Ruiz}, {Patel}, {Burrows}, {Mangano}, {Barthelmy},
  {Beardmore}, {Campana}, {Capalbi}, {Chincarini}, {Cusumano}, {Falcone},
  {Gehrels}, {Giommi}, {Goad}, {Godet}, {Hurkett}, {Kennea}, {Moretti},
  {O'Brien}, {Osborne}, {Romano}, {Tagliaferri}, \& {Wells}}]{Nousek:2006aa}
{Nousek}, J.~A., {Kouveliotou}, C., {Grupe}, D., {et~al.} 2006, \apj, 642, 389

\bibitem[{{Panaitescu}(2005{\natexlab{a}})}]{Panaitescu:2005aa}
{Panaitescu}, A. 2005{\natexlab{a}}, \mnras, 362, 921,
  \dodoi{10.1111/j.1365-2966.2005.09352.x}

\bibitem[{{Panaitescu}(2005{\natexlab{b}})}]{Panaitescu:2005ab}
---. 2005{\natexlab{b}}, \mnras, 363, 1409,
  \dodoi{10.1111/j.1365-2966.2005.09532.x}

\bibitem[{{Panaitescu} \& {Kumar}(2003)}]{Panaitescu:2003aa}
{Panaitescu}, A., \& {Kumar}, P. 2003, \apj, 592, 390, \dodoi{10.1086/375563}

\bibitem[{{Panaitescu} {et~al.}(1998){Panaitescu}, {M{\'e}sz{\'a}ros}, \&
  {Rees}}]{Panaitescu:1998aa}
{Panaitescu}, A., {M{\'e}sz{\'a}ros}, P., \& {Rees}, M.~J. 1998, \apj, 503,
  314, \dodoi{10.1086/305995}

\bibitem[{{Peng} {et~al.}(2005){Peng}, {K{\"o}nigl}, \& {Granot}}]{Peng:2005aa}
{Peng}, F., {K{\"o}nigl}, A., \& {Granot}, J. 2005, \apj, 626, 966,
  \dodoi{10.1086/430045}

\bibitem[{{Piro} {et~al.}(2005){Piro}, {De Pasquale}, {Soffitta}, {Lazzati},
  {Amati}, {Costa}, {Feroci}, {Frontera}, {Guidorzi}, {in't Zand}, {Montanari},
  \& {Nicastro}}]{Piro:2005aa}
{Piro}, L., {De Pasquale}, M., {Soffitta}, P., {et~al.} 2005, \apj, 623, 314,
  \dodoi{10.1086/428377}

\bibitem[{{Piro} {et~al.}(2019){Piro}, {Troja}, {Zhang}, {Ryan}, {van Eerten},
  {Ricci}, {Wieringa}, {Tiengo}, {Butler}, {Cenko}, {Fox}, {Khandrika},
  {Novara}, {Rossi}, \& {Sakamoto}}]{Piro:2019aa}
{Piro}, L., {Troja}, E., {Zhang}, B., {et~al.} 2019, \mnras, 483, 1912,
  \dodoi{10.1093/mnras/sty3047}

\bibitem[{{Planck Collaboration} {et~al.}(2016){Planck Collaboration}, {Ade},
  {Aghanim}, {Arnaud}, {Ashdown}, {Aumont}, {Baccigalupi}, {Banday},
  {Barreiro}, {Bartlett}, \& et~al.}]{Planck-Collaboration:2016aa}
{Planck Collaboration}, {Ade}, P.~A.~R., {Aghanim}, N., {et~al.} 2016, \aap,
  594, A13, \dodoi{10.1051/0004-6361/201525830}

\bibitem[{{Press} {et~al.}(2007){Press}, {Teukolsky}, {Vetterling}, \&
  {Flannery}}]{Press:2007aa}
{Press}, W.~H., {Teukolsky}, S.~A., {Vetterling}, W.~T., \& {Flannery}, B.~P.
  2007, Numerical Recipes 3rd Edition: The Art of Scientific Computing, 3rd
  edn. (New York, NY, USA: Cambridge University Press)

\bibitem[{{Racusin} {et~al.}(2009){Racusin}, {Liang}, {Burrows}, {Falcone},
  {Sakamoto}, {Zhang}, {Zhang}, {Evans}, \& {Osborne}}]{Racusin:2009aa}
{Racusin}, J.~L., {Liang}, E.~W., {Burrows}, D.~N., {et~al.} 2009, \apj, 698,
  43, \dodoi{10.1088/0004-637X/698/1/43}

\bibitem[{{Rees} \& {M{\'e}sz{\'a}ros}(1998)}]{Rees:1998aa}
{Rees}, M.~J., \& {M{\'e}sz{\'a}ros}, P. 1998, \apjl, 496, L1,
  \dodoi{10.1086/311244}

\bibitem[{{Rhoads}(1999)}]{Rhoads:1999aa}
{Rhoads}, J.~E. 1999, \apj, 525, 737, \dodoi{10.1086/307907}

\bibitem[{{Riess} {et~al.}(2016){Riess}, {Macri}, {Hoffmann}, {Scolnic},
  {Casertano}, {Filippenko}, {Tucker}, {Reid}, {Jones}, {Silverman},
  {Chornock}, {Challis}, {Yuan}, {Brown}, \& {Foley}}]{Riess:2016aa}
{Riess}, A.~G., {Macri}, L.~M., {Hoffmann}, S.~L., {et~al.} 2016, \apj, 826,
  56, \dodoi{10.3847/0004-637X/826/1/56}

\bibitem[{{Rossi} {et~al.}(2002){Rossi}, {Lazzati}, \& {Rees}}]{Rossi:2002aa}
{Rossi}, E., {Lazzati}, D., \& {Rees}, M.~J. 2002, \mnras, 332, 945,
  \dodoi{10.1046/j.1365-8711.2002.05363.x}

\bibitem[{{Rossi} {et~al.}(2004){Rossi}, {Lazzati}, {Salmonson}, \&
  {Ghisellini}}]{Rossi:2004aa}
{Rossi}, E.~M., {Lazzati}, D., {Salmonson}, J.~D., \& {Ghisellini}, G. 2004,
  \mnras, 354, 86, \dodoi{10.1111/j.1365-2966.2004.08165.x}

\bibitem[{{Ryan} {et~al.}(2015){Ryan}, {van Eerten}, {MacFadyen}, \&
  {Zhang}}]{Ryan:2015aa}
{Ryan}, G., {van Eerten}, H., {MacFadyen}, A., \& {Zhang}, B.-B. 2015, \apj,
  799, 3, \dodoi{10.1088/0004-637X/799/1/3}

\bibitem[{{Salmonson}(2003)}]{Salmonson:2003aa}
{Salmonson}, J.~D. 2003, \apj, 592, 1002, \dodoi{10.1086/375580}

\bibitem[{{Sari} \& {M{\'e}sz{\'a}ros}(2000)}]{Sari:2000aa}
{Sari}, R., \& {M{\'e}sz{\'a}ros}, P. 2000, \apjl, 535, L33,
  \dodoi{10.1086/312689}

\bibitem[{{Sari} {et~al.}(1999){Sari}, {Piran}, \& {Halpern}}]{Sari:1999aa}
{Sari}, R., {Piran}, T., \& {Halpern}, J.~P. 1999, \apjl, 519, L17,
  \dodoi{10.1086/312109}

\bibitem[{{Troja} {et~al.}(2016){Troja}, {Sakamoto}, {Cenko}, {Lien},
  {Gehrels}, {Castro-Tirado}, {Ricci}, {Capone}, {Toy}, {Kutyrev}, {Kawai},
  {Cucchiara}, {Fruchter}, {Gorosabel}, {Jeong}, {Levan}, {Perley},
  {Sanchez-Ramirez}, {Tanvir}, \& {Veilleux}}]{Troja:2016aa}
{Troja}, E., {Sakamoto}, T., {Cenko}, S.~B., {et~al.} 2016, \apj, 827, 102,
  \dodoi{10.3847/0004-637X/827/2/102}

\bibitem[{{Troja} {et~al.}(2017){Troja}, {Piro}, {van Eerten}, {Wollaeger},
  {Im}, {Fox}, {Butler}, {Cenko}, {Sakamoto}, {Fryer}, {Ricci}, {Lien}, {Ryan},
  {Korobkin}, {Lee}, {Burgess}, {Lee}, {Watson}, {Choi}, {Covino}, {D'Avanzo},
  {Fontes}, {Gonz{\'a}lez}, {Khandrika}, {Kim}, {Kim}, {Lee}, {Lee}, {Kutyrev},
  {Lim}, {S{\'a}nchez-Ram{\'{\i}}rez}, {Veilleux}, {Wieringa}, \&
  {Yoon}}]{Troja:2017aa}
{Troja}, E., {Piro}, L., {van Eerten}, H., {et~al.} 2017, \nat, 551, 71,
  \dodoi{10.1038/nature24290}

\bibitem[{{Troja} {et~al.}(2018{\natexlab{a}}){Troja}, {Piro}, {Ryan}, {van
  Eerten}, {Ricci}, {Wieringa}, {Lotti}, {Sakamoto}, \& {Cenko}}]{Troja:2018aa}
{Troja}, E., {Piro}, L., {Ryan}, G., {et~al.} 2018{\natexlab{a}}, \mnras, 478,
  L18, \dodoi{10.1093/mnrasl/sly061}

\bibitem[{{Troja} {et~al.}(2018{\natexlab{b}}){Troja}, {Ryan}, {Piro}, {van
  Eerten}, {Cenko}, {Yoon}, {Lee}, {Im}, {Sakamoto}, {Gatkine}, {Kutyrev}, \&
  {Veilleux}}]{Troja:2018ab}
{Troja}, E., {Ryan}, G., {Piro}, L., {et~al.} 2018{\natexlab{b}}, Nature
  Communications, 9, 4089, \dodoi{10.1038/s41467-018-06558-7}

\bibitem[{{Troja} {et~al.}(2019{\natexlab{a}}){Troja}, {van Eerten}, {Ryan},
  {Ricci}, {Burgess}, {Wieringa}, {Piro}, {Cenko}, \&
  {Sakamoto}}]{Troja:2019ab}
{Troja}, E., {van Eerten}, H., {Ryan}, G., {et~al.} 2019{\natexlab{a}}, \mnras,
  2169, \dodoi{10.1093/mnras/stz2248}

\bibitem[{{Troja} {et~al.}(2019{\natexlab{b}}){Troja}, {Castro-Tirado},
  {Becerra Gonzalez}, {Hu}, {Ryan}, {Cenko}, {Ricci}, {Novara},
  {Sanchez-Ramirez}, {Acosta-Pulido}, {Caballero Garcia}, {Guziy}, {Jeong},
  {Lien}, {Marquez}, {Pandey}, {Park}, {Tello}, {Sakamoto}, {Sokolov},
  {Sokolov}, {Tiengo}, {Valeev}, {Zhang}, \& {Veilleux}}]{Troja:2019aa}
{Troja}, E., {Castro-Tirado}, A.~J., {Becerra Gonzalez}, J., {et~al.}
  2019{\natexlab{b}}, arXiv e-prints, arXiv:1905.01290.
\newblock \doarXiv{1905.01290}

\bibitem[{{van Eerten}(2013)}]{van-Eerten:2013ab}
{van Eerten}, H. 2013, arXiv e-prints, arXiv:1309.3869.
\newblock \doarXiv{1309.3869}

\bibitem[{{van Eerten}(2018)}]{van-Eerten:2018ab}
---. 2018, International Journal of Modern Physics D, 27, 1842002,
  \dodoi{10.1142/S0218271818420026}

\bibitem[{{van Eerten} {et~al.}(2012){van Eerten}, {van der Horst}, \&
  {MacFadyen}}]{van-Eerten:2012ab}
{van Eerten}, H., {van der Horst}, A., \& {MacFadyen}, A. 2012, \apj, 749, 44,
  \dodoi{10.1088/0004-637X/749/1/44}

\bibitem[{{van Eerten} {et~al.}(2010){van Eerten}, {Zhang}, \&
  {MacFadyen}}]{van-Eerten:2010aa}
{van Eerten}, H., {Zhang}, W., \& {MacFadyen}, A. 2010, \apj, 722, 235,
  \dodoi{10.1088/0004-637X/722/1/235}

\bibitem[{{van Eerten} \& {MacFadyen}(2012{\natexlab{a}})}]{van-Eerten:2012aa}
{van Eerten}, H.~J., \& {MacFadyen}, A.~I. 2012{\natexlab{a}}, \apj, 751, 155,
  \dodoi{10.1088/0004-637X/751/2/155}

\bibitem[{{van Eerten} \& {MacFadyen}(2012{\natexlab{b}})}]{van-Eerten:2012ac}
---. 2012{\natexlab{b}}, \apjl, 747, L30, \dodoi{10.1088/2041-8205/747/2/L30}

\bibitem[{{Wu} \& {MacFadyen}(2018)}]{Wu:2018aa}
{Wu}, Y., \& {MacFadyen}, A. 2018, \apj, 869, 55,
  \dodoi{10.3847/1538-4357/aae9de}

\bibitem[{{Xie} {et~al.}(2018){Xie}, {Zrake}, \& {MacFadyen}}]{Xie:2018aa}
{Xie}, X., {Zrake}, J., \& {MacFadyen}, A. 2018, \apj, 863, 58,
  \dodoi{10.3847/1538-4357/aacf9c}

\bibitem[{{Yamazaki} {et~al.}(2002){Yamazaki}, {Ioka}, \&
  {Nakamura}}]{Yamazaki:2002aa}
{Yamazaki}, R., {Ioka}, K., \& {Nakamura}, T. 2002, \apjl, 571, L31,
  \dodoi{10.1086/341225}

\bibitem[{{Yamazaki} {et~al.}(2003){Yamazaki}, {Ioka}, \&
  {Nakamura}}]{Yamazaki:2003aa}
---. 2003, \apj, 593, 941, \dodoi{10.1086/376677}

\bibitem[{{Zhang} {et~al.}(2006){Zhang}, {Fan}, {Dyks}, {Kobayashi},
  {M{\'e}sz{\'a}ros}, {Burrows}, {Nousek}, \& {Gehrels}}]{Zhang:2006aa}
{Zhang}, B., {Fan}, Y.~Z., {Dyks}, J., {et~al.} 2006, \apj, 642, 354,
  \dodoi{10.1086/500723}

\bibitem[{{Zhang} \& {M{\'e}sz{\'a}ros}(2002)}]{Zhang:2002aa}
{Zhang}, B., \& {M{\'e}sz{\'a}ros}, P. 2002, \apj, 571, 876,
  \dodoi{10.1086/339981}

\bibitem[{{Zhang} {et~al.}(2015){Zhang}, {van Eerten}, {Burrows}, {Ryan},
  {Evans}, {Racusin}, {Troja}, \& {MacFadyen}}]{Zhang:2015aa}
{Zhang}, B.-B., {van Eerten}, H., {Burrows}, D.~N., {et~al.} 2015, The
  Astrophysical Journal, 806, 15, \dodoi{10.1088/0004-637X/806/1/15}

\end{thebibliography}

\end{document}